\documentclass[letterpaper,twocolumn,10pt]{article}
\usepackage{zhanggroup}

\usepackage{tikz}
\usepackage{tcolorbox}
\usepackage{amsfonts}
\usepackage{xspace} 
\usepackage{amsmath}
\usepackage{mathrsfs}
\usepackage{amssymb}
\usepackage{amsmath}
\usepackage{amsthm}
\usepackage[sort&compress,numbers]{natbib}
\usepackage{subcaption}
\usepackage{booktabs}
\usepackage{balance}
\usepackage{xcolor}
\hypersetup{
  colorlinks,
  linkcolor={blue!60!green},
  citecolor={green!60!blue},
  urlcolor={orange!60!red}
}

\usepackage[hang,flushmargin]{footmisc}
\usepackage{titlesec}
\titlespacing*{\section}{0pt}{*1.5}{2pt}
\titlespacing*{\subsection}{0pt}{*1.5}{2pt}
\titlespacing*{\subsubsection}{0pt}{*1}{1pt}
\usepackage[absolute]{textpos}

\newcommand{\mypara}[1]{\smallskip\noindent\textbf{#1:}}
\newcommand{\model}{\mathcal{M}}

\newcommand{\dataset}{\mathcal{D}}
\newcommand{\vTrigger}{\mathcal{T}}
\newcommand{\bdFunction}{\mathcal{A}}
\newcommand{\dTrigger}{{t}}
\newcommand{\featurevec}{x}

\newcommand{\scale}{s}
\newcommand{\backdoor}{bd}
\newcommand{\dLocation}{\kappa}
\newcommand{\vLocation}{\mathcal{K}}

\newcommand{\noisyBD}{Random Backdoor}
\newcommand{\ban}{BaN}
\newcommand{\cban}{c-BaN}
\newcommand{\clean}{c}
\newcommand{\outputvec}{y}
\newcommand{\dLabel}{\ell}

\newcommand{\vLabel}{{\mathcal{L}}}
\newcommand{\loss}{\varphi}

\begin{document}

\begin{textblock}{12}(2,1)
\centering
To Appear in the 7th IEEE European Symposium on Security and Privacy, June 6-10, 2022.
\end{textblock}

\date{}

\title{\Large \bf Dynamic Backdoor Attacks Against Machine Learning Models}


\author{
{\rm Ahmed Salem\textsuperscript{1}\thanks{The first two authors made equal contributions.}}\ \ \ \ \
{\rm Rui Wen\textsuperscript{2}\textsuperscript{\textcolor{blue!60!green}{$\ast$}}}\ \ \
{\rm Michael Backes\textsuperscript{2}}\ \ \
{\rm Shiqing Ma\textsuperscript{3}}\ \ \
{\rm Yang Zhang\textsuperscript{2}}\ \ \
\\
\\
\textsuperscript{1}\textit{Microsoft Research}\ \ \
\textsuperscript{2}\textit{CISPA Helmholtz Center for Information Security}\ \ \ \textsuperscript{3}\textit{Rutgers University}
}

\maketitle

\begin{abstract}
Machine learning (ML) has made tremendous progress during the past decade and is being adopted in various critical real-world applications. 
However, recent research has shown that ML models are vulnerable to multiple security and privacy attacks. 
In particular, backdoor attacks against ML models have recently raised a lot of awareness. 
A successful backdoor attack can cause severe consequences, such as allowing an adversary to bypass critical authentication systems.

Current backdooring techniques rely on adding static triggers (with fixed patterns and locations) on ML model inputs which are prone to detection by the current backdoor detection mechanisms. 
In this paper, we propose the first class of dynamic backdooring techniques against deep neural networks (DNN), namely Random Backdoor, Backdoor Generating Network (BaN), and conditional Backdoor Generating Network (c-BaN). 
Triggers generated by our techniques can have random patterns and locations, which reduce the efficacy of the current backdoor detection mechanisms. 
In particular, BaN and c-BaN based on a novel generative network are the first two schemes that algorithmically generate triggers. 
Moreover, c-BaN is the first conditional backdooring technique that given a target label, it can generate a target-specific trigger. 
Both BaN and c-BaN are essentially a general framework which renders the adversary the flexibility for further customizing backdoor attacks.

We extensively evaluate our techniques on three benchmark datasets: MNIST, CelebA, and CIFAR-10. 
Our techniques achieve almost perfect attack performance on backdoored data with a negligible utility loss. 
We further show that our techniques can bypass current state-of-the-art defense mechanisms against backdoor attacks, including ABS, Februus, MNTD, Neural Cleanse, and STRIP.
\end{abstract}

\section{Introduction}
\label{section:introduction}

\begin{figure*}[!t]
\centering
\begin{subfigure}{1.8\columnwidth}
\includegraphics[width=\columnwidth]{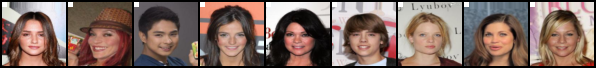}
\caption{Static backdoor}
\label{figure:badNets}
\end{subfigure}
\begin{subfigure}{1.8\columnwidth}
\includegraphics[width=\columnwidth]{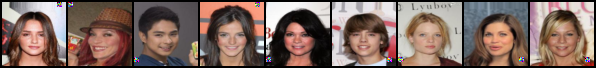}
\caption{Dynamic backdoor}
\label{figure:dynamic2}
\end{subfigure}
\caption{A comparison between static and dynamic backdoors. 
\autoref{figure:badNets} shows an example for static backdoors with a fixed trigger (white square at top left corner of the image). 
\autoref{figure:dynamic2} show examples for the dynamic backdoor with different triggers for the same target label. 
As the figures show, the dynamic backdoor trigger have different location and patterns, compared to the static backdoor where there is only a single trigger with a fixed location and pattern.}
\label{figure:example_multi_figure1}
\end{figure*} 

Machine learning (ML), represented by Deep Neural Network (DNN), has made tremendous progress during the past decade, and ML models have been adopted in a wide range of real-world applications including those that play critical roles. 
For instance, Apple's FaceID~\cite{FaceID} is using ML-based facial recognition systems for unlocking the mobile device and authenticating purchases in Apple Pay.
However, recent research has shown that machine learning models are vulnerable to various security and privacy attacks~\cite{PMJFCS16,PMGJCS17,XEQ18,SSSS17,SZHBFB19,TZJRR16,OASF18,WG18,LMALZWZ18,BNL12,JOBLNL18,SMKID18,YLZZ19,GDG17}.

In this work, we focus on backdoor attacks  against DNN models on image classification tasks, which are among the most successful ML applications deployed in the real world.
In the backdoor attack setting, an adversary trains an ML model which can intentionally misclassify any input with an added \emph{trigger} (a secret pattern constructed from a set of neighboring pixels, e.g., a white square) to a specific \emph{target label}. 
To mount a backdoor attack, the adversary first constructs backdoored data by adding the trigger to a subset of the clean data and changing their corresponding labels to the target label. 
Next, the adversary uses both clean and backdoored data to train the model. 
The clean and backdoored data are needed so the model can learn its original task and the backdoor behavior, simultaneously.
Backdoor attacks can cause severe security and privacy consequences.
For instance, an adversary can implant a backdoor in an authentication system to grant themselves unauthorized access.

Current state-of-the-art backdoor attacks~\cite{GDG17,LMALZWZ18,YLZZ19} generate static triggers, in terms of fixed trigger pattern and location (on the input).
For instance, \autoref{figure:badNets} shows an example of triggers constructed by Badnets~\cite{GDG17}, one of the most popular backdoor attack methods. 
As we can see, Badnets in this case uses a white square as a trigger and always places it in the top-left corner of all inputs. 
This static nature of triggers -with respect to their patterns and locations- is leveraged to create most of the current defenses against the backdoor attack~\cite{WYSLVZZ19,LLTMAZ19}.
Moreover, it facilitates linking backdoored data together, since they all have the same trigger at the same location.
This can result in efficiently patching the backdoored model without the need for any defense technique, if the defender gets access to a single backdoored input, i.e., the defender can extract the trigger from the backdoored input and use it to create a dataset to fine-tune and patch the backdoored model.

\subsection{Our Contributions}

In this work, we propose the first class of backdooring techniques against deep neural networks (DNN) models that generate dynamic triggers, in terms of trigger \emph{pattern} and \emph{location}.
We refer to our techniques as \emph{dynamic backdoor attacks}.
Dynamic backdoor attacks offer the adversary more flexibility, as they allow triggers to have different patterns and locations.
Moreover, our techniques largely reduce the efficacy of the current defense mechanisms demonstrated by our empirical evaluation.
\autoref{figure:dynamic2} shows an example of our dynamic backdoor attacks implemented in a model trained on the CelebA dataset~\cite{LLWT15}.
In addition, we extend our techniques to work for all labels of the backdoored ML model, while the current backdoor attacks only focus on a single or a few target labels.
This further increases the difficulty of our backdoors being mitigated.

In total, we propose 3 different dynamic backdoor techniques, namely, \emph{\noisyBD}, \emph{Backdoor Generating Network ({\ban})}, and \emph{conditional Backdoor Generating Network ({\cban})}.
In particular, the latter two attacks algorithmically generate triggers to mount backdoor attacks which are first of their kind.
In the following, we abstractly introduce each of our techniques.

\mypara{Random Backdoor}
In this approach, we construct triggers by sampling them from a uniform distribution.
Then, we place each randomly generated trigger at a random location for each input, which is then mixed with clean data to train the backdoor model.

\mypara{Backdoor Generating Network ({\ban})}
In our second technique, we propose a generative ML model, i.e., {\ban}, to generate triggers.
To the best of our knowledge, this is the first backdoor attack which uses a generative network to automatically construct triggers, which increases the flexibility of the adversary to perform backdoor attacks.
{\ban} is trained jointly with the backdoor model, it takes a latent code sampled from a uniform distribution to generate a trigger, then place it at a random location on the input, thus making the trigger dynamic in terms of pattern and location.
Moreover, {\ban} is essentially a general framework under which the adversary can change and adapt its loss function to their requirements.
For instance, if there is a specific backdoor defense in place, the adversary can evade the defense by adding a tailored discriminative loss in {\ban}.

\mypara{conditional Backdoor Generating Network ({\cban})}
Both of our {\noisyBD} and the {\ban} techniques can implement a dynamic backdoor for either a single target label or multiple target labels.
However, for the case of the multiple target labels, both techniques require each target label to have its unique trigger locations.
In other words, a single location cannot have triggers for different target labels.

Our last and most advanced technique overcomes the previous two techniques' limitation of having disjoint location sets for the multiple target labels.
In this technique, we transform the {\ban} into a conditional {\ban} ({\cban}), to force it to generate label specific triggers.
More specifically, we modify the {\ban}'s architecture to include the target label as an input, to generate a trigger for this specific label.
This target specific triggers property allows the triggers for different target labels to be positioned at any location.
In other words, each target label does not need to have its unique trigger locations.

To demonstrate the effectiveness of our proposed techniques, we perform empirical analysis with three ML model architectures over three benchmark datasets.
All of our techniques achieve almost a perfect backdoor accuracy, i.e., the accuracy of the backdoored model on the backdoored data is approximately 100\%, with a negligible utility loss.
For instance, our {\ban} trained models on CelebA~\cite{LLWT15} and MNIST~\cite{MNIST} datasets achieve 70\% and 99\% accuracy, respectively, which is the same accuracy as the clean models.
Also, {\cban}, {\ban}, and \noisyBD ~trained models achieve 92\%, 92.1\%, and 92\% accuracy 
on the CIFAR-10~\cite{CIFAR} dataset, respectively, which is almost the same as the performance of a clean model (92.4\%).
Moreover, we evaluate our techniques against three of the current state-of-the-art backdoor defense techniques, namely Neural Cleanse~\cite{WYSLVZZ19}, ABS~\cite{LLTMAZ19}, and STRIP~\cite{GXWCRN19}.
Our results show that our techniques can bypass these defenses.

In general, our contributions can be summarized as follows:
\begin{itemize}
    \item We broaden the class of backdoor attacks against deep neural networks (DNN) models by introducing the dynamic backdoor attacks.
    \item We propose both Backdoor Generating Network ({\ban}) and conditional Backdoor Generating Network ({\cban}), which are the first algorithmic backdoor paradigm.
    \item Our dynamic backdoor attacks achieve strong performance, while bypassing the current state-of-the-art backdoor defense techniques.
\end{itemize}

\section{Preliminaries}
\label{section:preliminaries}

In this section, we first introduce the machine learning classification setting.
Then we formalize backdoor attacks against ML models, and finally, we discuss the threat model we consider throughout the paper.

\subsection{Machine Learning Classification}

A machine learning classification model $\model$ is essentially a function that maps a feature vector $\featurevec$ from the feature space $\mathcal{X}$ to an output vector $\outputvec$ from the output space $\mathcal{Y}$, i.e., $\model(\featurevec)=\outputvec$.
Each entry $\outputvec_i$ in the vector $\outputvec$, corresponds to the posterior probability of the input vector $\featurevec$ being affiliated with the label $\dLabel_i\in \vLabel$, where $\vLabel$ is the set of all possible labels.
In this work, instead of $\outputvec$, we only consider the output of $\model$ as the label with the highest probability, i.e., $\model(x)=\text{argmax}_{\dLabel_i} \outputvec$.
To train $\model$, we need a dataset $\dataset$ which consists of pairs of labels and features vectors, i.e., $\dataset = \{(\featurevec_i,\dLabel_i)\}_{i \in \mathcal{N}}$ with $\mathcal{N}$ being the size of the dataset, and adopt some optimization algorithm, such as Adam, to learn the parameters of $\model$ following a defined loss function.

\subsection{Backdoor in Machine Learning Models}
\label{section:backdoors}

Intuitively, a backdoor in the ML settings resembles a hidden behavior of the model, which only happens when it is queried with an input containing a secret trigger.
This hidden behavior is usually the misclassification of an input feature vector to a desired target label.

A backdoored model $\model_{\backdoor}$ is expected to learn the mapping from feature vectors with triggers 
to their corresponding target label, i.e., any input with the trigger $\dTrigger_i$ should have the label $\dLabel_i$ as its output. 
To train such a model, an adversary needs both clean $\dataset_{\clean}$ (to preserve the model's utility) and backdoored data $\dataset_{\backdoor}$ (to implement the backdoor behaviour), where $\dataset_{\backdoor}$ is constructed by adding triggers on a subset of $\dataset_{\clean}$. 

Current backdoor attacks construct backdoors with static triggers, in terms of fixed trigger's pattern and location.
In this work, we introduce dynamic backdoors, where the trigger's pattern and location are dynamic.
In other words, a dynamic backdoor should have triggers with different values (pattern) and can be placed at different positions on the input (location).

A backdoor in an ML model is associated with a set of triggers $\vTrigger$, set of target labels $\vLabel^\prime$, and a backdoor adding function $\bdFunction$.
We first define the backdoor adding function $\bdFunction$ as: $\bdFunction(\featurevec,\dTrigger_i,\dLocation) =  \featurevec_{\backdoor}$, where $\featurevec$ is the input vector, $\dTrigger_i \in \vTrigger$ is the trigger, $\dLocation$ is the desired location to add the backdoor, and $\featurevec_{\backdoor}$ is the input vector $\featurevec$ with the backdoor inserted at the location $\dLocation$.  
More formally, $\bdFunction(\featurevec,\dTrigger_i,\dLocation) = \dTrigger_i \cdot \dLocation + \featurevec \cdot (1-\dLocation)$, where $\dLocation$ is a binary mask with ones at the specified location of the trigger.

Compared to the static backdoor attacks, dynamic backdoor attacks introduce new features for triggers, which give the adversary more flexibility and increase the difficulty of detecting such backdoors.
Namely, dynamic backdoors introduce different locations and patterns for the backdoor triggers.
These multiple patterns and locations for the triggers harden the detection of such backdoors, since the current design of defenses assumes a static behavior of backdoors.
Moreover, these triggers can be algorithmically generated, as will be shown later in \autoref{section:ban} and \autoref{section:cban}, which allows the adversary to customize the generated triggers.

\subsection{Threat Model}

As previously mentioned, the dynamic backdoor attacks are training time attacks, i.e., the adversary interferes with the training of the target model.
To implement our attacks, we assume the adversary controls the training of the target model and has access to the training data, following previous works on backdoor attacks~\cite{GDG17}.
We further relax this assumption (in~\autoref{section:relax}) to only assume the ability to poison the target model's training data.

To launch the attack -after publishing the model-, the adversary first adds a trigger to the input and then uses it to query the backdoored model.
This can happen either digitally, where the adversary digitally adds the trigger to the image, or physically, where the adversary prints the trigger and places it on the image similar to previous works~\cite{GDG17}.
This added trigger makes the backdoored model misclassify the input to the target label.
In practice, this can allow an adversary to bypass authentication systems to achieve their goal.

\section{Dynamic Backdoors}
\label{section:methodologies}

In this section, we propose three different techniques  for performing dynamic backdoor attacks, 
namely, {\noisyBD}, Backdoor Generating Network ({\ban}), and conditional Backdoor Generating Network ({\cban}).

\subsection{{\noisyBD}}
\label{section:noisyBD}

We start with our simplest approach, i.e., the {\noisyBD} technique.
Abstractly, the {\noisyBD} technique constructs triggers by sampling them from a uniform distribution, and adding them to the inputs at random locations.
We first introduce how to use our {{\noisyBD}} technique to implement a dynamic backdoor for a single target label, then we generalize it to consider multiple target labels.  

\mypara{Single Target Label}
We start with the simple case of considering dynamic backdoors for a single target label.
Intuitively, we construct a set of triggers ($\vTrigger$) and a set of possible locations ($\vLocation$), such that for any trigger sampled from $\vTrigger$ and added to any input at a random location sampled from $\vLocation$, the model will output the specified target label.
More formally, for any location $\dLocation_i \in \vLocation$, any trigger $\dTrigger_i \in \vTrigger$, and any input $\featurevec_i \in \mathcal{X}$:
\[
\model_{\backdoor}(\bdFunction(\featurevec_i,\dTrigger_i,\dLocation_i))=\dLabel
\]
where $\dLabel$ is the target label, $\vTrigger$ is the set of triggers, and $\vLocation$ is the set of locations.

To implement such a backdoor in a model, an adversary needs first to select their desired trigger locations, and create the set of possible locations $\vLocation$.
Then, they use both clean and backdoored data to update the model for each epoch.
More concretely, the adversary trains the model as mentioned in~\autoref{section:backdoors} with the following two differences.

\begin{enumerate}
    \item First, instead of using a fixed trigger for all inputs, each time the adversary wants to add a trigger to an input, they sample a new trigger from a uniform distribution, i.e., $\dTrigger \sim \mathcal{U}(0,1)$. 
    Here, the set of possible triggers $\vTrigger$ contains the full range of all possible values for the triggers, since the trigger is randomly sampled from a uniform distribution.
    \item Second, instead of placing the trigger in a fixed location, they place it at a random location $\dLocation$, sampled from the predefined set of location, i.e., $\dLocation \in \vLocation$.
\end{enumerate}

This technique is not only limited to the uniform distribution, but the adversary can use different distributions like the Gaussian distribution to construct triggers.
Using different distributions can, for example, help the adversary to change the appearance of the used triggers.

Finally, the adversary does not need access to the training of the target model for this technique.
Instead, they can backdoor a target model by only adding the backdoored data to its training set, i.e., poison the training set.

\begin{figure}[!t]
\centering
\includegraphics[width=0.4\columnwidth]{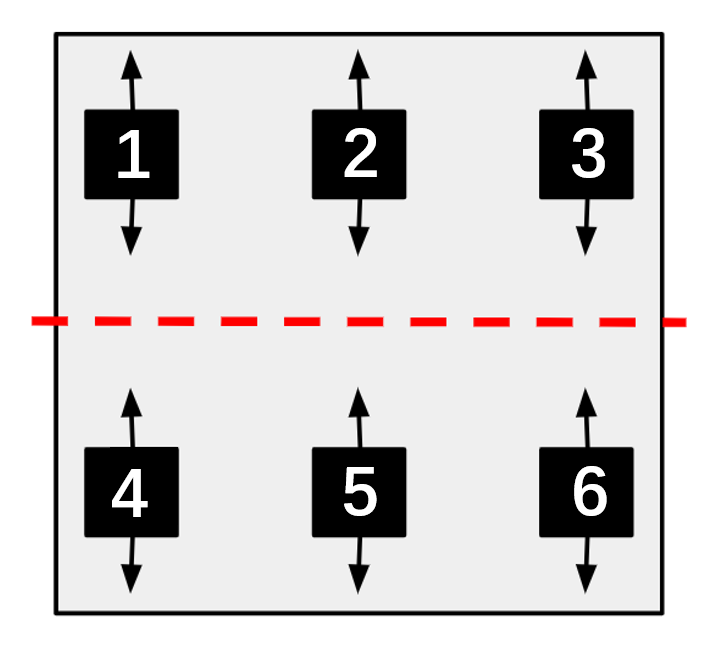}
\caption{An illustration of our location setting technique for 6 target labels. 
The red dotted line demonstrates the boundary of the vertical movement for each target label. }
\label{figure:locationsNoise}
\end{figure} 

\mypara{Multiple Target Labels}
Next, we consider the more complex case of having multiple target labels.
Without loss of generality, we consider implementing a backdoor for each label in the dataset, since this is the most challenging setting. 
However, our techniques can be applied to any smaller subset of labels.
This means that for any label $\dLabel_i \in \vLabel$, there exists a trigger $\dTrigger$ which when added to the input $\featurevec$ at a location $\dLocation$, will make the model $\model_{\backdoor}$ output $\dLabel_i$.
More formally,
\[
\forall \dLabel_i \in \vLabel\ \exists\ \dTrigger,\dLocation: \model_{\backdoor}(\bdFunction(\featurevec,\dTrigger,\dLocation))=\dLabel_i
\]

To achieve the dynamic backdoor behaviour in this setting, each target label should have a set of possible triggers and a set of possible locations.
More formally,
\[
\forall \dLabel_i \in \vLabel\ \exists\ \vTrigger_i,\vLocation_i
\]
where $\vTrigger_i$ is the set of possible triggers and $\vLocation_i$ is the set of possible locations for the target label $\dLabel_i$.

We generalize the {{\noisyBD}} technique by dividing the set of possible locations $\vLocation$ into disjoint subsets for each target label, while keeping the trigger construction method the same as in the single target label case, i.e., the triggers are still sampled from a uniform distribution.
For instance, for the target label $\dLabel_i$, we sample a set of possible locations $\vLocation_i$, where $\vLocation_i$ is a subset of $\vLocation$ ($\vLocation_i \subset \vLocation$).

The adversary can construct the disjoint sets of possible locations as follows:

\begin{enumerate}
    \item First, the adversary selects all possible triggers locations and constructs the set $\vLocation$.
    \item Second, for each target label $\dLabel_i$, they construct the set of possible locations for this label $\vLocation_i$ by sampling the set $\vLocation$. 
    Then, they remove the sampled locations from the set $\vLocation$.
\end{enumerate}

We propose the following simple algorithm to assign the locations for the different target labels.
However, an adversary can construct the location sets arbitrarily with the only restriction that no location can be used for more than one target label.

We uniformly split the image into non-intersecting regions, and assign a region for each target label, in which the triggers' locations can move vertically.
\autoref{figure:locationsNoise} shows an example of our location setting technique for a use case with 6 target labels.
As the figure shows, each target label has its own region, for example, label $1$ occupies the top left region of the image.
We stress that this is one way of dividing the location set $\vLocation$ to the different target labels.
However, an adversary can choose a different way of splitting the locations inside $\vLocation$ to the different target labels. 
The only requirement the adversary has to fulfill is to avoid assigning a location for different target labels.
Later, we will show how to overcome this limitation with our more advanced {\cban} technique. 

\begin{figure*}[!t]
\centering
\begin{subfigure}{1\columnwidth}
\includegraphics[width=1\columnwidth]{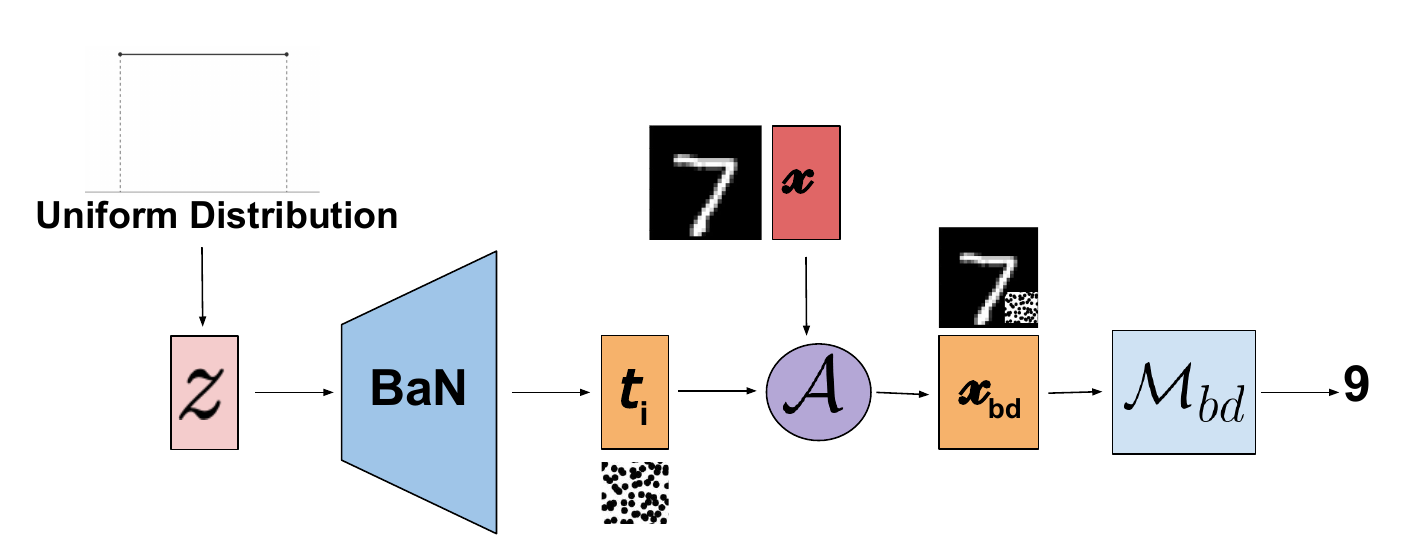}
\caption{{\ban}}
\label{figure:BaN}
\end{subfigure}
\begin{subfigure}{1\columnwidth}
\includegraphics[width=1\columnwidth]{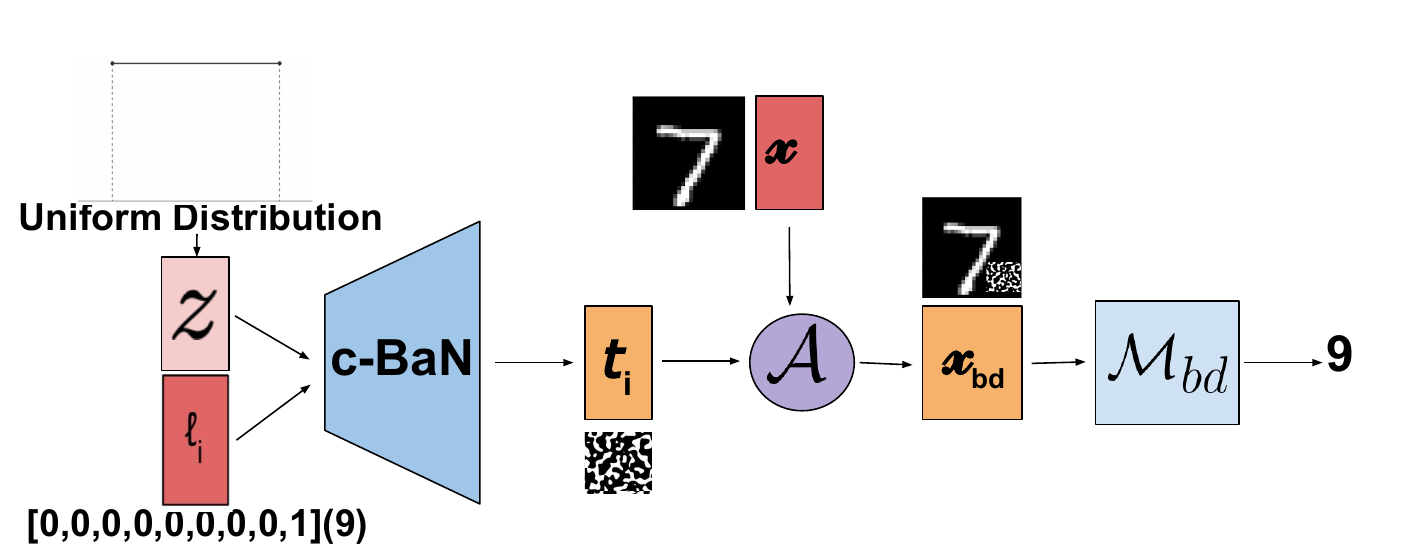}
\caption{{\cban}}
\label{figure:cBaN}
\end{subfigure}
\caption{An overview of the {\ban} and {\cban} techniques.}
\end{figure*} 

\subsection{Backdoor Generating Network ({\ban})}
\label{section:ban}

Our Random Backdoor technique successfully implements dynamic triggers, however, it offers the adversary limited flexibility as triggers are sampled from a preset distribution.
Moreover, the triggers are sampled independently of the target model.
In other words, the Random Backdoor technique does not search for the best triggers to implement the backdoor attack.
To address these limitations, we introduce our second technique to implement dynamic backdoors, namely, Backdoor Generating Network ({\ban}).
{\ban} is the first approach to algorithmically generate backdoor triggers, instead of using fixed triggers or sampling triggers from a uniform distribution (as in \autoref{section:noisyBD}).

{\ban} is inspired by the state-of-the-art generative models, i.e., Generative Adversarial Networks (GANs)~\cite{GPMXWOCB14}.
However, it is different from the original GANs in the following aspects.
First, instead of generating images, it generates backdoor triggers.
Second, we jointly train the {\ban}'s generator with the target model instead of the discriminator, to learn (the generator) and implement (the target model) the best patterns for the backdoor triggers.

After training, the {\ban} can generate a trigger ($\dTrigger$) for each noise vector ($z \sim \mathcal{U}(0,1)$).
This trigger is then added to an input using the backdoor adding function $\bdFunction$, to create the backdoored input as shown in \autoref{figure:BaN}.
Similar to the previous approach ({\noisyBD}), the generated triggers are placed at random locations.

In this section, we first introduce the {\ban} technique for a single target label, then we generalize it for multiple target labels.

\mypara{Single Target Label}
We start with presenting how to implement a dynamic backdoor for a single target label, using our {\ban} technique.
First, the adversary creates the set $\vLocation$ of the possible locations.
They then jointly train the {\ban} with the backdoored $\model_{\backdoor}$ model as follows:

\begin{enumerate}
    \item The adversary starts each training epoch by querying the clean data to the backdoored model $\model_{\backdoor}$. 
    Then, they calculate the clean loss $\loss_{\clean}$ between the ground truth and the output labels. 
    We use the cross-entropy loss for our clean loss, which is defined as follows:
    \[
      \sum\limits_{i}  \outputvec_{i} \log(\hat{\outputvec}_{i})
    \]
    where $\outputvec_{i}$ is the true probability of label $\dLabel_i$ and $\hat{\outputvec}_{i}$ is our predicted probability of label $\dLabel_i$. 
    \item They then generate $n$ noise vectors, where $n$ is the batch size.
    \item On the input of the $n$ noise vectors, the {\ban} generates $n$ triggers.
    \item The adversary then creates the backdoored data by adding the generated triggers to the clean data using the backdoor adding function $\bdFunction$.
    \item They then query the backdoored data to the backdoored model $\model_{\backdoor}$ and calculates the backdoor loss $\loss_{\backdoor}$ on the model's output and the target label. 
    Similar to the clean loss, we use the cross-entropy loss as our loss function for $\loss_{\backdoor}$.
    \item Finally, the adversary updates the backdoor model $\model_{\backdoor}$ using both the clean and backdoor losses ($\loss_{\clean} + \loss_{\backdoor}$) and updates the {\ban} with the backdoor loss ($\loss_{\backdoor}$).
\end{enumerate}

We show later in \autoref{section:relax} how to simplify the threat model for the {\ban} technique to only assume the ability to poison the training data, i.e., the adversary backdoors the target model without interfering with its training algorithm.

\mypara{Multiple Target Labels}
We now consider the more complex case of building a dynamic backdoor for multiple target labels using our {\ban} technique.
To recap, our {\ban} generates general triggers and does not label specific triggers.
In other words, the same trigger pattern can be used to trigger multiple target labels.
Thus similar to the {\noisyBD}, we depend on the location of the triggers to determine the output label.

We follow the same approach of the {\noisyBD} technique to assign different locations for different target labels (\autoref{section:noisyBD}), to generalize the {\ban} technique.
More concretely, the adversary implements the dynamic backdoor for multiple target labels using the {\ban} technique as follows:

\begin{enumerate}
    \item The adversary starts by creating disjoint sets of locations for all target labels.    
    \item Next, they follow the same steps as in training the backdoor for a single target label, while repeating from step 2 to 5 for each target label and adding all their backdoor losses together. 
    More formally, for the multiple target label case the backdoor loss is defined as:
    $\sum_i^{|\vLabel'|}\loss_{\backdoor_i}$, where $\vLabel'$ is the set of target labels, and $\loss_{\backdoor_i}$ is the backdoor loss for target label $\dLabel_i$.
\end{enumerate}

\subsection{conditional Backdoor Generating Network ({\cban})}
\label{section:cban}

So far, we have proposed two techniques to implement dynamic backdoors for both single and multiple target labels, i.e, {\noisyBD} (\autoref{section:noisyBD}) and {\ban} (\autoref{section:ban}).
To recap, both techniques have the limitation of not having label specific triggers and only depending on the trigger location to determine the target label.
We now introduce our third and most advanced technique, the conditional Backdoor Generating Network ({\cban}), which overcomes this limitation.
More concretely, with the {\cban} technique any location $\dLocation$ inside the location set $\vLocation$ can be used to trigger any target label.
To achieve this location independency, the triggers need to be label specific.
Therefore, we convert the Backdoor Generating Network ({\ban}) into a conditional Backdoor Generating Network ({\cban}).
More specifically, we add the target label as an additional input to the {\ban} for conditioning it to generate target specific triggers.

We construct {\cban} by adding an additional input layer to {\ban} to include the target label as an input.
\autoref{figure:cBaN} represents an illustration of {\cban}.
As the figure shows, the noise vector and the target label are encoded to latent vectors with the same size (to give equal weights for both inputs).
These two latent vectors are then concatenated and used as an input to the next layer.

The {\cban} is trained similarly to the {\ban}, with the following two exceptions.

\begin{enumerate}
	\item First, the adversary does not have to create disjoint sets of locations for all target labels (step 1), they can use the complete location set $\vLocation$ for all target labels.
	\item Second, instead of using only the noise vectors as an input to the {\ban}, the adversary one-hot encodes the target label, then use it together with the noise vectors as the input to the {\cban}. 
\end{enumerate}

Similar to {\ban}, we later (\autoref{section:relax}) show how to simplify the threat model for the {\cban}.

To use the {\cban}, the adversary first samples a noise vector and one-hot encodes the label.
Then, they input both of them to the {\cban}, which generates a trigger.
The adversary uses the backdoor adding function $\bdFunction$ to add the trigger to the target input.
Finally, they query the backdoored input to the backdoored model, which will output the target label.
We visualize the complete pipeline of using the {\cban} technique in \autoref{figure:cBaN}.

In this section, we have introduced three techniques for implementing dynamic backdoors, namely, the {\noisyBD}, the Backdoor Generating Network ({\ban}), and the conditional Backdoor Generating Network ({\cban}).
These three dynamic backdoor techniques present a framework to generate dynamic backdoors for different settings.
For instance, our framework can generate target specific triggers' pattern using the {\cban}, or target specific triggers' location like the {\noisyBD} and {\ban}.
More interestingly, our framework allows the adversary to customize their backdoor by adapting the backdoor loss functions.
For instance, the adversary can adapt to different defenses against the backdoor attack that can be modeled as a machine learning model.
This can be achieved by adding any defense as a discriminator into the training of the {\ban} or {\cban}.
Adding this discriminator will penalize/guide the backdoored model to bypass the modeled defense.

\section{Evaluation}
\label{section:evaluation}

In this section, we first introduce our datasets and experimental settings.
Next, we evaluate all of our three techniques, i.e., {\noisyBD}, Backdoor Generating Network ({\ban}), and conditional Backdoor Generating Network ({\cban}).
We then evaluate our three dynamic backdoor techniques against the current state-of-the-art backdoor defense techniques, and study the effect of different hyperparameters on their performance.
Finally, we demonstrate how to relax the threat model and propose new defenses against dynamic backdoor attacks.

\subsection{Datasets Description}

We utilize three image datasets to evaluate our techniques, including MNIST, CelebA, and CIFAR-10.
We use these three datasets since they are widely used as benchmark datasets for various security/privacy and computer vision tasks, however, our attack can be easily generalized to other datasets with different types of data (by adapting the architectures of the {\ban} and {\cban}).
We briefly describe each of them below.

\mypara{MNIST}
The MNIST dataset~\cite{MNIST} is a 10-class dataset consisting of $70,000$ grey-scale $28\times28$ images.
Each of these images contains a handwritten digit in its center.
The MNIST dataset is a balanced dataset, i.e, each class is represented with $7,000$ images.

\mypara{CIFAR-10}
The CIFAR-10 dataset~\cite{CIFAR} is composed of $60,000$ $32\times32$ colored images which are equally distributed on the following $10$ classes: Airplane, automobile, bird, cat, deer, dog, frog, horse, ship, and truck.

\mypara{CelebA}
The CelebA dataset~\cite{LLWT15} is a large-scale face attributes dataset with more than $200$K colored celebrity images, each annotated with $40$ binary attributes.
We select the top three most balanced attributes including Heavy Makeup, Mouth Slightly Open, and Smiling.
Then we concatenate them into $8$ classes to create a multiple label classification task.
For our experiments, we scale the images to $64\times64$ and randomly sample $10,000$ images for training, and another $10,000$ for testing.
Finally, it is important to mention that unlike the MNIST and CIFAR-10 datasets, this dataset is highly imbalanced.

\subsection{Experimental Setup}

We first present our target models, then the evaluation metrics. 
For the target models' architecture, we use the VGG-19~\cite{SZ15} for the CIFAR-10 dataset, and build our own convolution neural networks (CNN) for the CelebA and MNIST datasets.
More concretely, we use 3 convolution layers and 5 fully connected layers for the CelebA CNN.
And 2 convolution layers and 2 fully connected layers for the MNIST CNN.  
Moreover, we use dropout for both the CelebA and MNIST models to avoid overfitting.

For {\ban}, we use the following architecture: 
\begin{tcolorbox}[boxsep=1pt,left=2pt,right=2pt,top=0.5 pt,bottom=0pt]
\emph{Backdoor Generating Network ({\ban})'s architecture:}
\begin{align*}
 z \rightarrow
\texttt{FullyConnected(64)} & \\
\texttt{FullyConnected(128)} & \\
\texttt{FullyConnected(128)} & \\
\texttt{FullyConnected(|\dTrigger|)}&\\
\texttt{Sigmoid} & \rightarrow \dTrigger
\end{align*}
\end{tcolorbox}
\noindent 
Here, \texttt{FullyConnected($x$)} denotes a fully connected layer with $x$ hidden units, $|\dTrigger|$ denotes the size of the required trigger, and \texttt{Sigmoid} is the Sigmoid function.
We adopt ReLU as the activation function for all layers, and apply dropout after all layers except the first and last ones.

For {\cban}, we use the following architecture: 
\begin{tcolorbox}[boxsep=1pt,left=2pt,right=2pt,top=0.5 pt,bottom=0pt]
\emph{conditional Backdoor Generating Network ({\cban})'s architecture:}
\begin{align*}
 z,\dLabel \rightarrow
2 \times \texttt{FullyConnected(64)} & \\
\texttt{FullyConnected(128)} & \\
\texttt{FullyConnected(128)} & \\
\texttt{FullyConnected(128)} & \\
\texttt{FullyConnected(|\dTrigger|)}&\\
\texttt{Sigmoid} & \rightarrow \dTrigger
\end{align*}
\end{tcolorbox}
\noindent 
The first layer consists of two separate fully connected layers, where each one of them takes an independent input, i.e., the first takes the noise vector $z$ and the second takes the target label $\dLabel$.
The outputs of these two layers are then concatenated and used as an input to the next layer (see~\autoref{section:cban}).
Similar to {\ban}, we adopt ReLU as the activation function for all layers and apply dropout after all layers except the first and last one. 

For evaluating the dynamic backdoor attacks' performance, we define the following two metrics:
\emph{Backdoor success rate} which calculates the backdoored model's accuracy on the backdoored data;
\emph{Model utility} which measures the original functionality of the backdoored model.
We quantify the model utility by comparing the accuracy of the backdoored model with the accuracy of a clean model on clean data.
Closer accuracies imply a better model utility.
All of our experiments are implemented using Pytorch and our code will be published for reproducibility.

\subsection{\noisyBD}
\label{section:noisyBdRes}

We now evaluate the performance of our first dynamic backdooring technique, namely, the {\noisyBD}.
We use all three datasets for the evaluation. 
First, we evaluate the single target label case, where we only implement a backdoor for a single target label.
Then we evaluate the more generalized case, i.e., the multiple target labels case, where we implement a backdoor for all possible labels in the dataset.

For both the single and multiple target label cases, we split each dataset into training and testing datasets.
The training dataset is used to train the MNIST and CelebA models from scratch.
For CIFAR-10, we use a pre-trained VGG-19 model.
For evaluating our models, we use the testing dataset as our clean testing dataset.
And construct a backdoored testing dataset, by adding triggers to all members of the testing dataset.
To recap, for the {\noisyBD} technique, we construct the triggers by sampling them from uniform distribution, and add them to the images using the backdoor adding function $\bdFunction$.
We use the backdoored testing dataset to calculate the backdoor success rate, and the training dataset to train a clean model -for each dataset- to evaluate the backdoored model's ($\model_{\backdoor}$) utility.

\begin{figure}[!t]
\centering
\includegraphics[width=0.8\columnwidth]{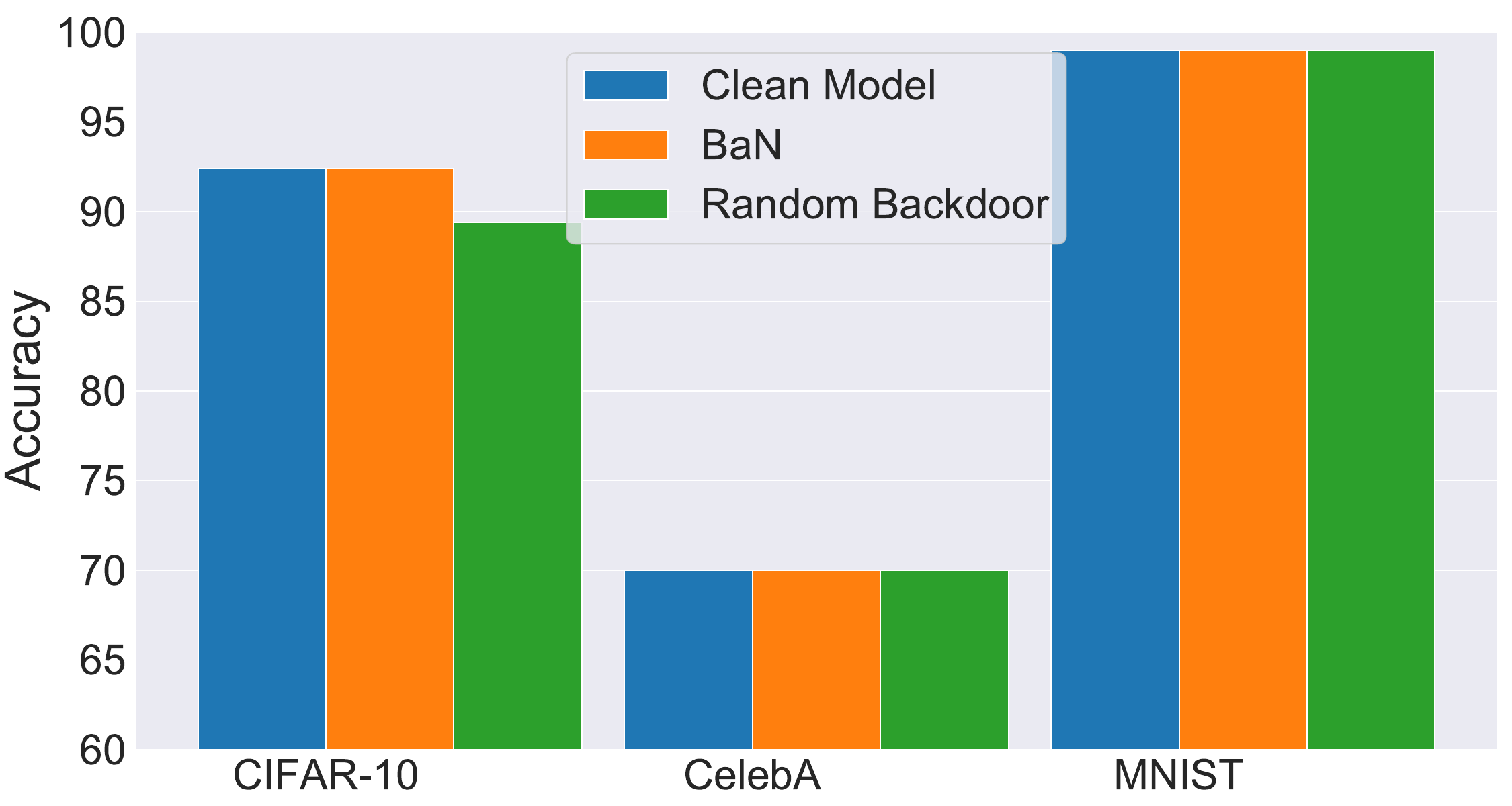}
\caption{The result of our dynamic backdoor techniques for a single target label on the clean testing dataset.}
\label{figure:singleLabel}
\end{figure} 

We follow~\autoref{section:noisyBD} to train our backdoored model $\model_{\backdoor}$ for both the single and multiple target labels cases.
Abstractly, for each epoch, we update the backdoored model $\model_{\backdoor}$ using both the clean and backdoor losses $\loss_{\clean} + \loss_{\backdoor}$.
For the set of possible locations $\vLocation$, we use four possible locations.

The backdoor success rate is always 100\% for both the single and multiple target labels cases on all three datasets, hence, we only focus on the backdoored model's ($\model_{\backdoor}$) utility.

\mypara{Single Target Label}
We first present our results for the single target label case.
\autoref{figure:singleLabel} compares the accuracies of the backdoored model $\model_{\backdoor}$ and the clean model $\model$.
As the figure shows, our backdoored models achieve the same performance as the clean models for both the MNIST and CelebA datasets, i.e., 99\% for MNIST and 70\% for CelebA.
For the CIFAR-10 dataset, there is a slight drop in performance, which is less than 2\%.
This shows that our {\noisyBD} technique can implement a perfectly functioning backdoor, i.e., the backdoor success rate of $\model_{\backdoor}$ is 100\% on the backdoored testing dataset, with a negligible utility loss.

\begin{figure}[!t]
\centering
\begin{subfigure}{0.8\columnwidth}
\includegraphics[width=\columnwidth]{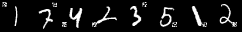}
\caption{\noisyBD}
\label{figure:singleLabelNoise}
\end{subfigure}
\begin{subfigure}{0.8\columnwidth}
\includegraphics[width=\columnwidth]{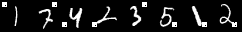}
\caption{{\ban}}
\label{figure:singleLabelBaN}
\end{subfigure}
\begin{subfigure}{0.8\columnwidth}
\includegraphics[width=\columnwidth]{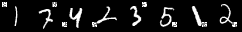}
\caption{{\ban} with higher randomness}
\label{figure:singleLabelBaNMoreRandom}
\end{subfigure}
\caption{The result of our \noisyBD~(\autoref{figure:singleLabelNoise}), {\ban} (\autoref{figure:singleLabelBaN}), and {\ban}  with higher randomness (\autoref{figure:singleLabelBaNMoreRandom}) techniques for a single target label (0).}
\label{figure:example_multi_figure}
\end{figure} 

To visualize the output of our {\noisyBD} technique, we first randomly sample 8 images from the MNIST dataset, and then use the {\noisyBD} technique to construct triggers for them.
Finally, we add these triggers to the images using the backdoor adding function $\bdFunction$, and show the result in~\autoref{figure:singleLabelNoise}.
As the figure shows, the triggers all look distinctly different and are located at different locations as expected.

\mypara{Multiple Target Labels}
Second, we present our results for the multiple target label case.
To recap, we consider all possible labels for this case.
For instance, for the MNIST dataset, we consider all digits from 0 to 9 as our target labels.
We train our {\noisyBD} models for the multiple target labels as mentioned in~\autoref{section:noisyBD}.

We use a similar evaluation setting to the single target label case, with the following exception.
To evaluate the performance of the backdoored model $\model_{\backdoor}$ with multiple target labels, we construct a backdoored testing dataset for each target label by generating and adding triggers to the clean testing dataset.
In other words, we use all images in the testing dataset to evaluate all possible labels.

Similar to the single target label case, we focus on the accuracy on the clean testing dataset, since the backdoor success rate for all models on the backdoored testing datasets are approximately 100\% for all target labels.

\begin{figure}[!t]
\centering
\includegraphics[width=0.8\columnwidth]{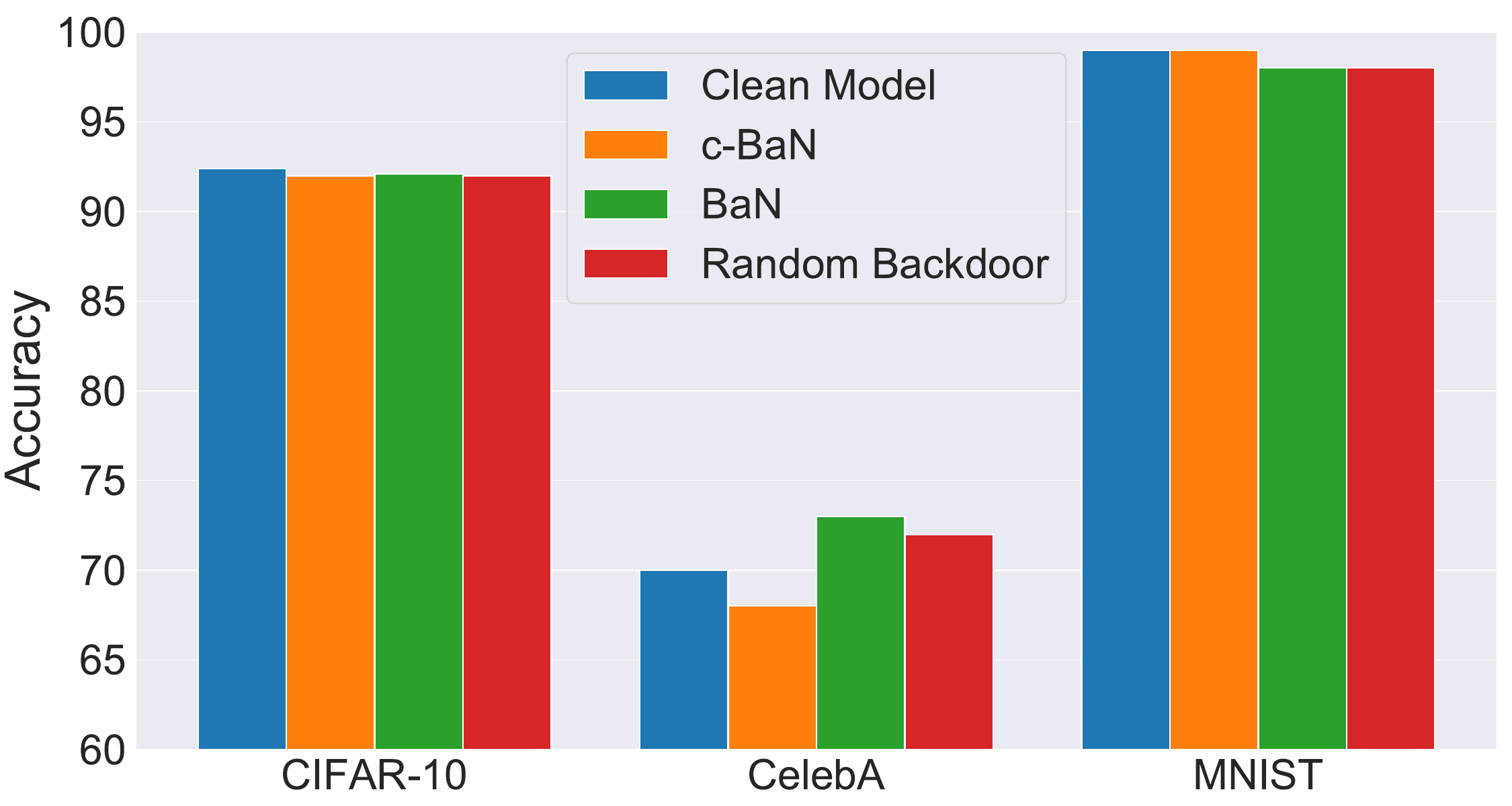}
\caption{The result of our dynamic backdoor techniques for multiple target label on the clean testing dataset.}
\label{figure:multiLabel}
\end{figure}

We use the clean testing datasets to evaluate the backdoored model's $\model_{\backdoor}$ utility, i.e., we compare the performance of the backdoored model $\model_{\backdoor}$ with the clean model $\model$ in \autoref{figure:multiLabel}.
As the figure shows, using our {\noisyBD} technique, we are able to train backdoored models that achieve similar performance as the clean models for all datasets.
For instance, for the CIFAR-10 dataset, our {\noisyBD} technique achieves 92\% accuracy, which is very similar to the accuracy of the clean model (92.4\%).
For the CelebA dataset, the {\noisyBD} technique achieves a slightly (about 2\%) better performance than the clean model.
We believe this is due to the regularization effect of the {\noisyBD} technique.
Finally, for the MNIST dataset, both models achieve a similar performance with just 1\% difference between the clean model (99\%) and the backdoored one (98\%).

\begin{figure*}[!t]
\centering
\begin{subfigure}{1.8\columnwidth}
\includegraphics[width=\columnwidth]{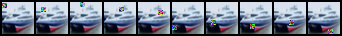}
\caption{\noisyBD}
\label{figure:AllLabelNoise}
\end{subfigure}
\begin{subfigure}{1.8\columnwidth}
\includegraphics[width=\columnwidth]{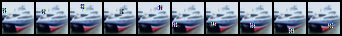}
\caption{{\ban}}
\label{figure:AllLabelBaN}
\end{subfigure}
\begin{subfigure}{1.8\columnwidth}
\includegraphics[width=\columnwidth]{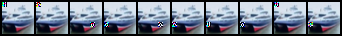}
\caption{{\cban}}
\label{figure:AllLabelcBaN}
\end{subfigure}
\caption{The visualization result of our \noisyBD~ (\autoref{figure:AllLabelNoise}), {\ban} (\autoref{figure:AllLabelBaN}), and {\cban} (\autoref{figure:AllLabelcBaN}) techniques for all labels of the CIFAR-10 dataset.}
\label{figure:allLabels}
\end{figure*}

To visualize the output of our {\noisyBD} technique on multiple target labels, we construct triggers for all possible labels in the CIFAR-10 dataset, and use $\bdFunction$ to add them to a randomly sampled image from the CIFAR-10 clean testing dataset.
\autoref{figure:AllLabelNoise} shows the image with different triggers.
The different patterns and locations used for the different target labels can be clearly demonstrated in~\autoref{figure:AllLabelNoise}.
For instance, comparing the location of the trigger for the first and sixth images, the triggers are in the same horizontal position but a different vertical position, as previously illustrated in~\autoref{figure:locationsNoise}.

Moreover, we further visualize in~\autoref{figure:Label5Noise} the dynamic behavior of the triggers generated by our {\noisyBD} technique.
Without loss of generality, we generate triggers for the target label 5 (plane) and add them to randomly sampled CIFAR-10 images.
To make it clear, we train the backdoor model $\model_{\backdoor}$ for all possible labels set as target labels, but we visualize the triggers for a single label to show the dynamic behaviour of our {\noisyBD} technique with respect to the triggers' pattern and locations.
As \autoref{figure:Label5Noise} shows, the generated triggers have different patterns and locations for the same target label, which achieves our desired dynamic behavior.

\subsection{Backdoor Generating Network ({\ban})}
\label{section:banEval}

Next, we evaluate our {\ban} technique.
We follow the same evaluation settings for the {\noisyBD} technique, except with respect to how the triggers are generated.
We train our {\ban} model and generate the triggers as mentioned in~\autoref{section:ban}.

\mypara{Single Target Label} 
Similar to the {\noisyBD}, the {\ban} technique achieves a perfect backdoor success rate with a negligible utility loss.
\autoref{figure:singleLabel} compares the performance of the backdoored models -trained using the {\ban} technique- with the clean models, when tested using the clean testing dataset.
As \autoref{figure:singleLabel} shows, our {\ban} trained backdoored models achieve 99\%, 92.4\% and 70\% accuracy on the MNIST, CIFAR-10, and CelebA datasets, respectively, which is the same performance of the clean models.

We visualize the {\ban} generated triggers using the MNIST dataset in \autoref{figure:singleLabelBaN}.
To construct the figure, we use the {\ban} to generate multiple triggers -for the target label 0-, then we add them on a set of randomly sampled MNIST images using the backdoor adding function $\bdFunction$.

The generated triggers look very similar as shown in~\autoref{figure:singleLabelBaN}.
This behaviour is expected as the MNIST dataset is simple, and the {\ban} technique does not have any explicit loss to enforce the network to generate different triggers.
However, to show the flexibility of our approach, we increase the randomness of the {\ban} network by simply adding one more dropout layer after the last layer, to avoid the overfitting of the {\ban} model to a unique pattern.
We show the results of the {\ban} model with higher randomness in \autoref{figure:singleLabelBaNMoreRandom}.
The resulting model still achieves the same performance, i.e., 99\% accuracy on the clean data and 100\% backdoor success rate, but as the figure shows the triggers look significantly different.
This again shows that our framework can easily adapt to the requirements of an adversary. 

These results together with the results of the {\noisyBD} (\autoref{section:noisyBdRes}) clearly show the effectiveness of both of our proposed techniques, for the single target label case.
They are both able to achieve almost the same accuracy of a clean model, with a 100\% working backdoor, for a single target label.

\mypara{Multiple Target Labels} 
Similar to the single target label case, we focus on the backdoored models' performance on the clean testing dataset, as our {\ban} backdoored models achieve a perfect accuracy on the backdoored testing dataset, i.e., the backdoor success rate for all datasets is approximately 100\% for all target labels.

We compare the performance of the {\ban} backdoored models with one of clean models using the clean testing dataset in \autoref{figure:multiLabel}.
Our {\ban} backdoored models are able to achieve almost the same accuracy as the clean model for all datasets, as can be shown in \autoref{figure:multiLabel}.
For instance, for the CIFAR-10 dataset, our {\ban} achieves 92.1\% accuracy, which is only 0.3\% less than the performance of the clean model (92.4\%).
Similar to the {\noisyBD} backdoored models, our {\ban} backdoored models achieve a marginally better performance for the CelebA dataset.
More concretely, our {\ban} backdoored models trained for the CelebA dataset achieve about 2\% better performance than the clean model, on the clean testing dataset.
We also believe this improvement is due to the regularization effect of the {\ban} technique.
Finally, for the MNIST dataset, our {\ban} backdoored models achieve strong performance on the clean testing dataset (98\%), which is just 1\% lower than the performance of the clean models (99\%).

Similar to the {\noisyBD}, we visualize the results of the {\ban} backdoored models with two figures.
The first (\autoref{figure:AllLabelBaN}) shows the different triggers for the different target labels on the same CIFAR-10 image, and the second (\autoref{figure:Label5BaN}) shows the different triggers for the same target label (plane) on randomly sampled CIFAR-10 images.
As both figures show, the {\ban} generated triggers achieves the dynamic behaviour in both locations and patterns.
For instance, for the same target label (\autoref{figure:Label5BaN}), the patterns of the triggers look significantly different and the locations vary vertically.
Similarly, for different target labels (\autoref{figure:AllLabelBaN}), both the pattern and location of triggers are significantly different.

\subsection{conditional Backdoor Generating Network ({\cban})}
\label{section:cBaNEval}

Next, we evaluate our conditional Backdoor Generating Network ({\cban}) technique. 
For the single target label case, the {\cban} technique is the same as the {\ban} technique.
Thus, we only consider the multiple target labels case in this section. 

We follow a similar setup as the one introduced in \autoref{section:banEval}, with the exception on how to train the backdoored model $\model_{\backdoor}$ and generate the triggers.
We follow \autoref{section:cban} to train the backdoored model and generate the triggers.
For the set of possible locations $\vLocation$, we use four possible locations.

We compare the performance of the {\cban} with the other two techniques in addition to the clean model.
All of our three dynamic backdoor techniques achieve an almost perfect backdoor success rate on the backdoored testing datasets, hence similar to the previous sections, we focus on the performance on the clean testing datasets. 

\autoref{figure:multiLabel} compares the accuracy of the backdoored and clean models using the clean testing dataset, for all of our three dynamic backdoor techniques.
As the figure shows, all of our dynamic backdoored models have similar performance as the clean models.
For instance, for the CIFAR-10 dataset, our {\cban}, {\ban} and {\noisyBD} achieve 92\%, 92.1\% and 92\% accuracy, respectively, which is very similar to the accuracy of the clean model (92.4\%).
Also for the MNIST dataset, all models achieve very similar performance with no difference between the clean and {\cban} models (99\%) and only 1\% difference between them, and the {\ban} and {\noisyBD} models (98\%).

Similar to the previous two techniques, we visualize the dynamic behaviour of the {\cban} backdoored models using two different figures. 
First, by generating triggers for all possible labels and adding them on a CIFAR-10 image in~\autoref{figure:AllLabelcBaN}.
More generally, \autoref{figure:allLabels} shows the visualization of all three dynamic backdoor techniques in the same settings, i.e., backdooring a single image to all possible labels.
As the figure shows, the \noisyBD~\autoref{figure:AllLabelNoise} has the most random patterns, which is expected as they are sampled from a uniform distribution.
The figure also shows the different triggers' patterns and locations used for the different techniques.
For instance, each target label in the {\noisyBD} (\autoref{figure:AllLabelNoise}) and {\ban} (\autoref{figure:AllLabelBaN}) techniques have a unique (horizontal) location, unlike the {\cban} (\autoref{figure:AllLabelcBaN}) generated triggers, which different target labels can share the same locations, as can be shown for example in the first, second, and ninth images.
To recap, both the {\noisyBD} and {\ban} techniques split the location set $\vLocation$ on all target labels, such that no two labels share a location, unlike the {\cban} technique which does not have this limitation.

\begin{figure}[!t]
\centering
\begin{subfigure}{0.7\columnwidth}
\includegraphics[width=\columnwidth]{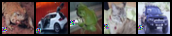}
\caption{\noisyBD}
\label{figure:Label5Noise}
\end{subfigure}
\begin{subfigure}{0.7\columnwidth}
\includegraphics[width=\columnwidth]{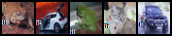}
\caption{{\ban}}
\label{figure:Label5BaN}
\end{subfigure}
\begin{subfigure}{0.7\columnwidth}
\includegraphics[width=\columnwidth]{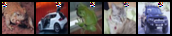}
\caption{{\cban}}
\label{figure:Label5cBaN}
\end{subfigure}
\caption{The result of our \noisyBD ~(\autoref{figure:Label5Noise}), {\ban} (\autoref{figure:Label5BaN}), and {\cban}  (\autoref{figure:Label5cBaN}) techniques for the target target label 5 (plane).}
\label{figure:label5}
\end{figure}

Second, we visualize the dynamic behaviour of our techniques, by generating triggers for the same target label 5 (plane) and adding them to a set of randomly sampled CIFAR-10 images.
\autoref{figure:label5} compares the visualization of our three different dynamic backdoor techniques in this setting.
More concretely, we train the backdoor model $\model_{\backdoor}$ for all possible labels set as target labels.
Then, for space restrictions, we plot the backdoored inputs for a single target label.
As the figure shows, the \noisyBD ~(\autoref{figure:Label5Noise}) and {\ban} (\autoref{figure:Label5BaN}) generated triggers can move vertically, however, they have a fixed position horizontally as mentioned in~\autoref{section:noisyBD} and illustrated in~\autoref{figure:locationsNoise}.
The {\cban} (\autoref{figure:Label5cBaN}) triggers also show different locations.
However, the locations of these triggers are more distant and can be shared for different target labels, unlike the other two techniques.
Furthermore, the figure shows that most of our triggers have different patterns for our techniques for the same target label, which achieves our targeted dynamic behavior concerning the patterns and locations of the triggers.

\begin{figure}[!t]
\centering
\begin{subfigure}{0.48\columnwidth}
\includegraphics[width=1\columnwidth]{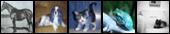}
\includegraphics[width=1\columnwidth]{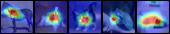}
\caption{Clean images}
\label{figure:cleanGRAD}
\end{subfigure}
\begin{subfigure}{0.48\columnwidth}
\includegraphics[width=1\columnwidth]{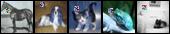}
\includegraphics[width=1\columnwidth]{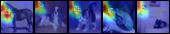}
\caption{{\noisyBD}}
\label{figure:noisyGRAD}
\end{subfigure}
\begin{subfigure}{0.48\columnwidth}
\includegraphics[width=1\columnwidth]{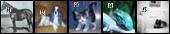}
\includegraphics[width=1\columnwidth]{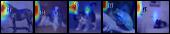}
\caption{{\ban}}
\label{figure:BaNGRAD}
\end{subfigure}
\begin{subfigure}{0.48\columnwidth}
\includegraphics[width=1\columnwidth]{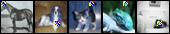}
\includegraphics[width=1\columnwidth]{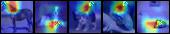}
\caption{{\cban}}
\label{figure:cBaNGRAD}
\end{subfigure}
\caption{Visualization of attention maps for all our techniques using the Grad-CAM technique.}
\label{figure:GRADCAM}
\end{figure} 

Finally, we compare the attention of the backdoored models on both clean and backdoored inputs.
We use the Gradient-weighted Class Activation Mapping (Grad-CAM) technique~\cite{SCDVPB17} to compute the attention maps for our backdoored models.
These maps show the most influential parts of the input that resulted in the model's output.
\autoref{figure:GRADCAM} depicts the results of our three different techniques.
As expected all backdoored models mainly focus on the triggers in backdoored inputs and the main objects in the clean ones.

\subsection{Evaluating Against Current State-Of-The-Art Defenses}

We now evaluate our attacks against the current state-of-the-art backdoor defenses.
Backdoor defenses can be classified into the following two categories, data-based defenses and model-based defenses.
On the one hand, data-based defenses focus on identifying if a given input is clean or contains a trigger.
On the other hand, model-based defenses focus on identifying if a given model is clean or backdoored.

We first evaluate our attacks against model-based defenses, then we evaluate them against data-based ones.

\mypara{Model-based Defense}
We evaluate all of our dynamic backdoor techniques in the multiple target label case against three of the current state-of-the-art model-based defenses, namely, Neural Cleanse~\cite{WYSLVZZ19}, ABS~\cite{LLTMAZ19}, and MNTD~\cite{XWLBGL21}.

We start by evaluating the ABS defense.
We use the CIFAR-10 dataset to evaluate this defense, since it is the only supported dataset by the published defense model.
As expected, running the ABS model against our dynamic backdoored ones does not result in detecting any backdoor for all of our models.

For Neural Cleanse, we use all three datasets to evaluate our techniques against it.
Similar to ABS, all of our models are predicted to be clean models.
Moreover, in multiple cases, our models had a lower anomaly index (the lower the better) than the clean model.

We believe that both of these defenses fail to detect our backdoors for two reasons.
First, we break one of their main assumption, i.e., that the triggers are static in terms of location and pattern.
Second, we implement a backdoor for all possible labels, which makes the detection a more challenging task.

Finally, we evaluate the MNTD defense.
To this end, we use the CIFAR-10 dataset to evaluate our three backdoor techniques.
Following the same setting in~\cite{JLG22}, we build 200 shadow benign and backdoored models to train 50 meta-classifiers sequentially for further evaluation.
The meta-classifier takes a target model as its input and outputs a score. This score represents the likelihood of the model being backdoored, i.e., a higher score indicates the target model is more likely to be backdoored.

Our results show that the score predicted by the MNTD meta classifier drops from 67.08($\pm$ 20.49) for static backdoors to 3.05($\pm$ 0.82), 0.54($\pm$ 0.83), and 1.47($\pm$ 0.87) for the random backdoor, BaN, and cBaN backdoors.
This significant reduction of scores (with at least a factor of 22$\times$) demonstrates the advantage of our techniques compared to the static ones.
In this setting, each meta-classifier would output a score, then based on a threshold, the decision if the model is backdoored or not is made~\cite{JLG22}.
We use the default threshold (median of training models' scores), which results in 98\%, 74\%, and 98\% accuracy for the random backdoor, BaN, and cBaN techniques, respectively.
This percentage corresponds to the number of meta-classifiers correctly classifying the model as a backdoored one.
To improve the stealthiness of our backdoored models, we add a discriminator when training the models, aiming to lower the score predicted by the MNTD meta classifier.
More concretely, we train a local meta-classifier (with a disjoint dataset compared to the one used for evaluation) and use it as our discriminator.
We demonstrate this with the cBaN technique; however, it can be easily extended to the other two techniques.
Using this technique, our results are significantly improved, i.e.,  only a single meta-classifier out of the 50 classified the model as a backdoored one.
In other words, the detection accuracy is dropped to 2\%, with a negligible performance drop, i.e., the ASR and utility dropped by less than 1\%.

This again demonstrates that our dynamic backdoor techniques are more stealthy than the static ones. 
Moreover, they can be easily adapted to bypass backdoor defenses, e.g., by adding the corresponding discriminator as mentioned in~\autoref{section:cban}.

\mypara{Data-based Defense}
Next, we evaluate some of the current state-of-the-art data-based defenses.
Namely, we start by evaluating STRIP~\cite{GXWCRN19}, then Februus~\cite{DAR20}.

STRIP tries to identify if a given input is clean or contains a trigger.
It works by creating multiple images from the input image by fusing it with multiple clean images one at a time.
Then STRIP applies all fused images to the target model and calculates the entropy of predicted labels.
Backdoored inputs tend to have lower entropy compared to the clean ones.

We use all of our three datasets to evaluate the {\cban} models against this defense.
First, we scale the patterns by half while training the backdoored models, to make them more susceptible to changes.
Second, for the MNIST dataset, we move the possible locations to the middle of the image to overlap with the image content, since the value of the MNIST images at the corners are always 0.
All trained scaled backdoored models achieve similar performance to the non-scaled backdoored models.

\begin{figure}[!t]
\centering
\begin{subfigure}{0.3\columnwidth}
\includegraphics[width=\columnwidth]{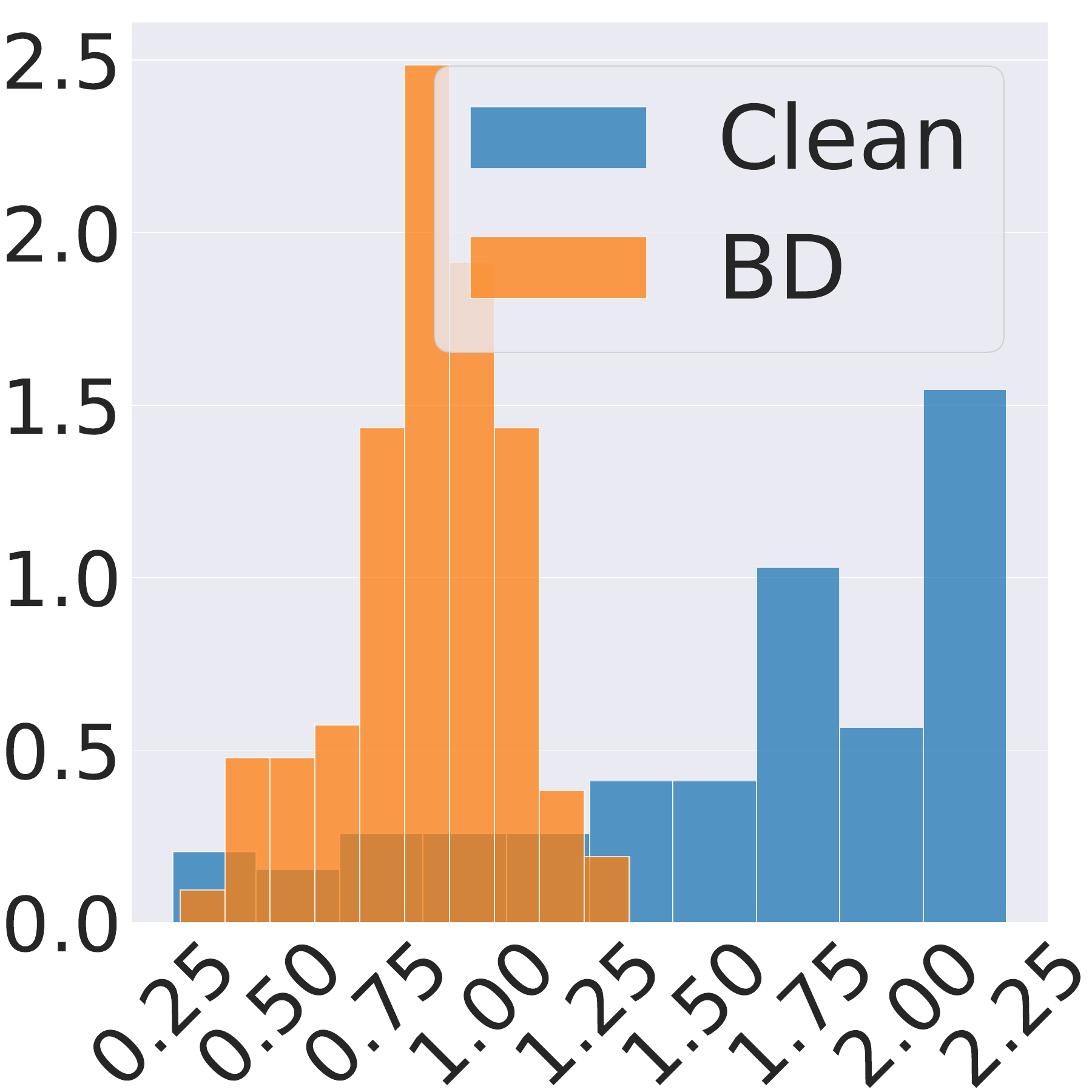}
\caption{CIFAR-10}
\label{figure:CIFAREntropy}
\end{subfigure}
\begin{subfigure}{0.3\columnwidth}
\includegraphics[width=\columnwidth]{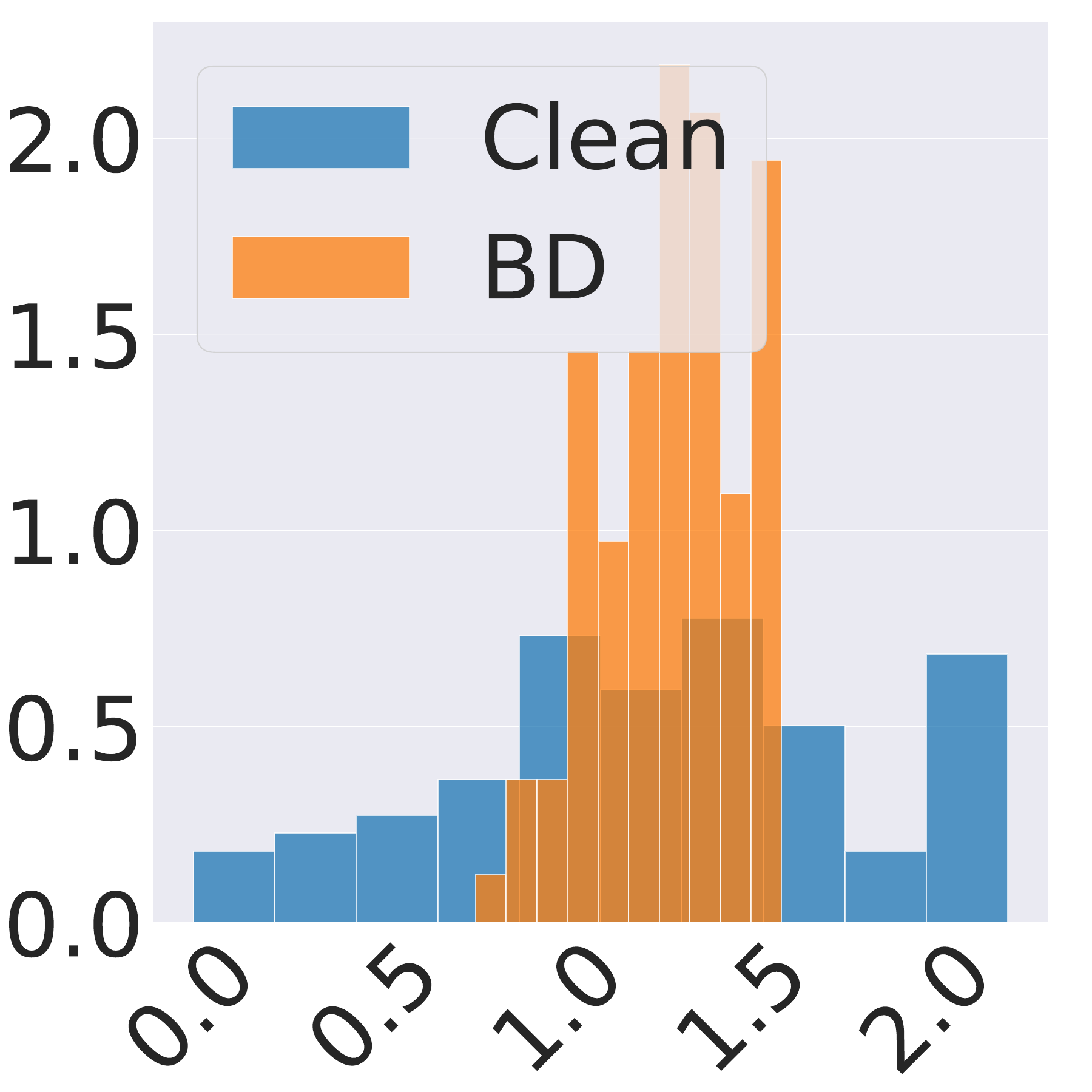}
\caption{MNIST}
\label{figure:MNISTEntropy}
\end{subfigure}
\begin{subfigure}{0.3\columnwidth}
\includegraphics[width=\columnwidth]{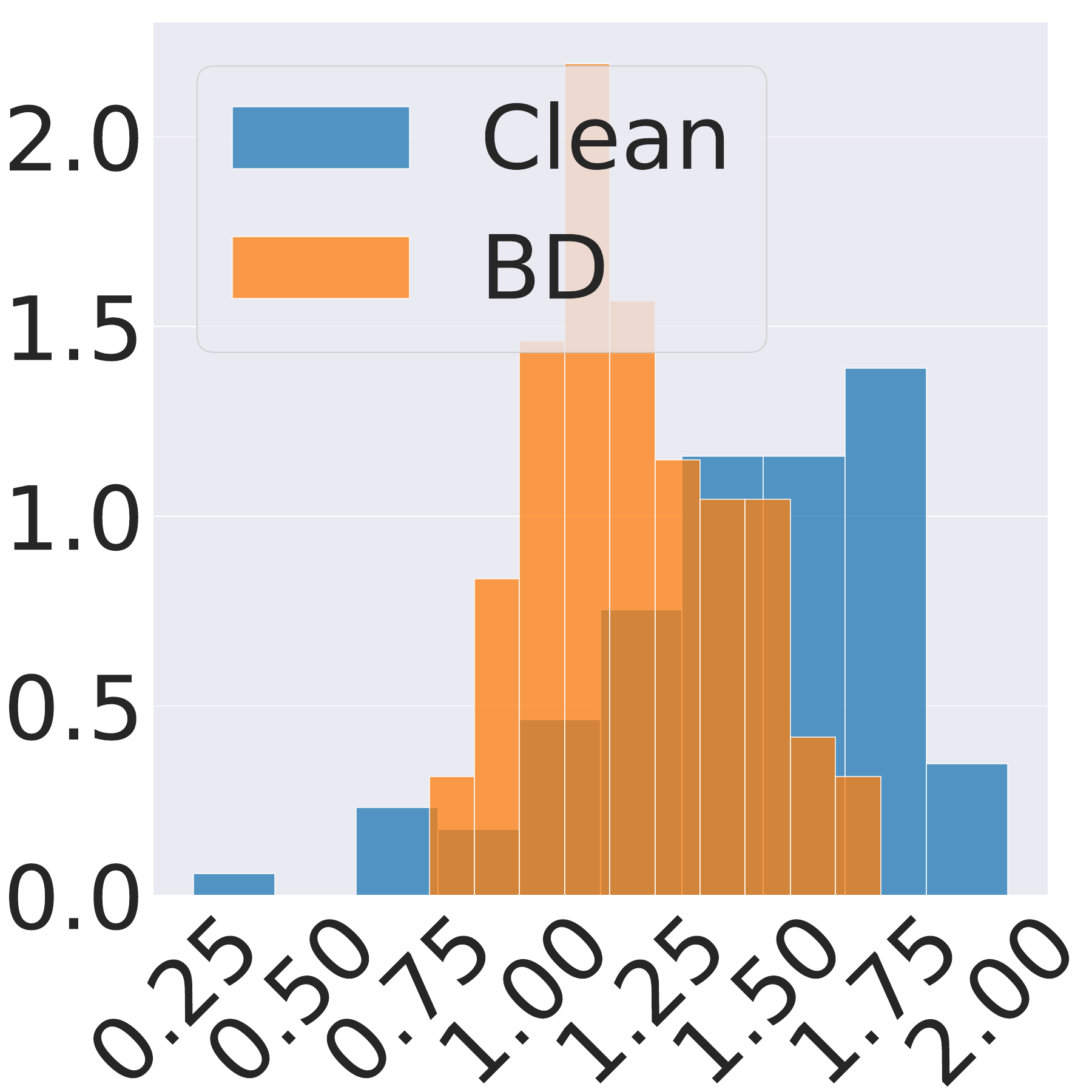}
\caption{CelebA}
\label{figure:celebAEntropy}
\end{subfigure}
\caption{The histogram of the entropy of the backdoored vs clean input, for our best performing labels against the STRIP defense, for the CIFAR-10 (\autoref{figure:CIFAREntropy}), MNIST (\autoref{figure:MNISTEntropy}), and CelebA  (\autoref{figure:celebAEntropy}) datasets.}
\label{figure:entropy}
\end{figure} 

Our backdoored models successfully flatten the distribution of entropy for the backdoored data, for a subset of target labels.
In other words, the distribution of entropy for our backdoored data overlaps with the distributions of entropy of the clean data.
This subset of target labels makes picking a threshold to identify backdoored data from clean data impossible without increasing the false positive rate, i.e., various clean images will be detected as backdoored ones. 
We visualize the entropy of our best performing labels against the STRIP defense in~\autoref{figure:entropy}.
Moreover, since our dynamic backdoors can generate dynamic triggers for the same input and target label, the adversary can keep querying the target model while backdooring the input with a freshly generated trigger until the model accepts it.

Next, we evaluate Februus.
Intuitively, Februus first detects the trigger from backdoored samples before removing it and patching the image.
To detect these triggers, Februus first uses GradCAM to identify the influential region on the input.
Then based on a security hyperparameter --which is dependent on the underlying task--, it decides if this area is to be removed and replaced by a neutral color.
Finally, Februus develops a GAN-based inpainting technique to restore the image before querying it to the target model.

As the training code of Februus is not public yet, we only use CIFAR-10 -- since it is the only dataset we consider that has its Februus models available -- to evaluate against our different backdoor techniques.

Our results show that Februus only succeeds in dropping the ASR of our random backdoor, BaN, and cBaN backdoored models from 100\% to approximately 80.5\%, 81.7\%, and 72\%, respectively.
This demonstrates the strong performance of our attack against the data-based defenses, especially compared to the static backdoored -- whose ASR drops to 0.25\% when applying Februus--.

These results against the data and model-based defenses show the effectiveness of our dynamic backdoor attacks, and opens the door for designing backdoor detection systems that work against both static and dynamic backdoors.

\subsection{Evaluating Different Hyperparameters}

We now evaluate the effect of different hyperparameters for our dynamic backdooring techniques.
We start by evaluating the percentage of the backdoored data needed to implement a dynamic backdoor into the model.
Then, we evaluate the effect of increasing the size of the location set $\vLocation$.
Finally, we evaluate the size of the trigger and the possibility of making it more transparent, i.e., instead of replacing the original values in the input with the backdoor, we fuse them.

\mypara{Proportion of the Backdoored Data}
We start by evaluating the percentage of backdoored data needed to implement a dynamic backdoor in the model.
We use the MNIST dataset and the {\cban} technique to perform the evaluation.
First, we construct different training datasets with different percentages of backdoored data.
More concretely, we try all proportions from 10\% to 50\%, with a step of 10.
In this setting, 10\% means that 10\% of the data is backdoored, and 90\% is clean.
Our results show that using 30\% is already enough to get a perfectly working dynamic backdoor, i.e., the model has a similar performance like a clean model on the clean dataset (99\% accuracy), and 100\% backdoor success rate on the backdoored dataset.
For any percentage below 30\%, the accuracy of the model on clean data is still the same, however, the performance on the backdoored dataset starts degrading. 
This demonstrates the ability of the adversary to implement dynamic backdoor attacks with $30\%$ overhead for each target label, compared to training a clean model.

\mypara{Number of Locations}
Second, we explore the effect of increasing the size of the set of possible locations ($\vLocation$) for the {\cban} technique.
We use the  CIFAR-10 dataset to train a backdoored model using the {\cban} technique, but with more than double the size of $\vLocation$, i.e., 8 locations.
The trained model achieves similar performance on the clean (92\%) and backdoored (100\%) datasets.
We then doubled the size again to have 16 possible locations in $\vLocation$, and the model again achieves the same results on both clean and backdoored datasets.
We repeat the experiment with the CelebA datasets and achieve similar results, i.e., the performance of the model with a larger set of possible locations is similar to the previously reported one.
However, when we try to completely remove the location set $\vLocation$ and consider all possible locations with a sliding window, the performance on both clean and backdoored datasets drops significantly.

\begin{figure}[!t]
\centering
\includegraphics[width=0.8\columnwidth]{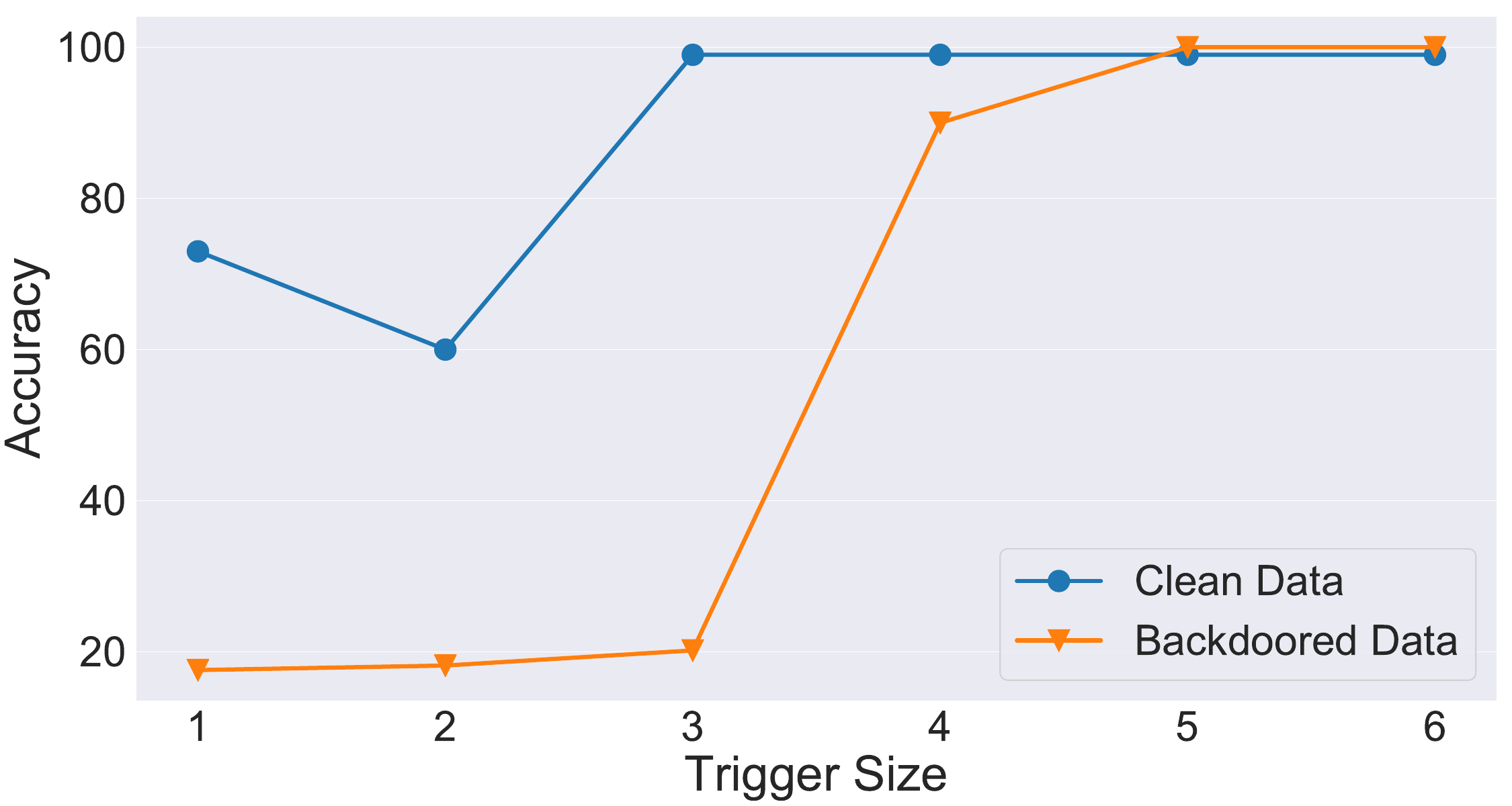}
\caption{The result of trying different trigger sizes for the {\cban} technique on the MNIST dataset. 
The figure shows for each trigger size the accuracy on the clean and backdoored testing datasets.}
\label{figure:triggerSize}
\end{figure} 

\mypara{Trigger Size}
Next, we evaluate the effect of the trigger size on our {\cban} technique using the MNIST dataset.
We train different models with the {\cban} technique, while setting the trigger size from 1 to 6.
We define the trigger size to be the width and height of the trigger.
For instance, a trigger size of 3 means that the trigger is $3\times3$ pixels.

We calculate the accuracy on the clean and backdoored testing datasets for each trigger size, and show our results in~\autoref{figure:triggerSize}.
Our results show that the smaller the trigger, the harder it is for the model to implement the backdoor behaviour.
Moreover, small triggers confuse the model, which results in reducing the model's utility.
As \autoref{figure:triggerSize} shows, a trigger with the size 5 achieves a perfect accuracy (100\%) on the backdoored testing dataset, while preserving the accuracy on the clean testing dataset (99\%).

\mypara{Transparency of the Triggers}
Finally, we evaluate the effect of making the trigger more transparent.
More specifically, we change the backdoor adding function $\bdFunction$ to apply a weighted sum, instead of replacing the original input's values.
Abstractly, we define the weighted sum of the trigger and the image as:
$\featurevec_{\backdoor} = \scale \cdot \dTrigger + (1-\scale) \cdot \featurevec$, where $\scale$ is the scale controlling the transparency rate, $\featurevec$ is the input and $\dTrigger$ is the trigger.
We implement this weighted sum only at the location of the trigger, while maintaining the remaining of the input unchanged.

\begin{figure}[!t]
\centering
\includegraphics[width=0.7\columnwidth]{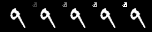}
\caption{An illustration of the effect of using different transparency scales (from 0 to 1 with step of 0.25) when adding the trigger. 
Scale 0 (the most left image) shows the original input, and scale 1 (the most right image) the original backdoored input without any transparency.}
\label{figure:trans}
\end{figure}

We use the MNIST dataset and {\cban} technique to evaluate the scale from 0 to 1, with a step of 0.25.
\autoref{figure:trans} visualizes the effect of varying the scale when adding a trigger to an input.

Our results show that our technique can achieve the same performance on both the clean (99\%) and backdoored (100\%) testing datasets, when setting the scale to 0.5 or higher.
However, when the scale is set below 0.5, the performance starts degrading on the backdoored dataset but stays the same on the clean dataset.
We repeat the same experiments for the CelebA and CIFAR-10 datasets and find similar results.
 
\begin{figure}
\centering
\includegraphics[width=0.8\columnwidth]{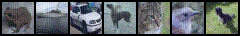}
\caption{Visualization of the {\cban} backdoored images when setting the transparency scale to 0.1.}
\label{figure:transparentCIFAR}
\end{figure}

We believe that the transparency of our triggers can be further increased when using triggers with larger sizes.
To this end, we use the CIFAR-10 dataset to repeat the experiments previously mentioned in this section.
However, we set the trigger size to be the size of the image.
Our experiments show that in this setting, our dynamic backdoor attacks can still achieve a perfect attack success rate (100\%) with a negligible drop in utility (0.3\%) when setting the scale to $0.1$.
More concretely, the model's accuracy on clean data is 91.7\% compared to the 92\% accuracy of the backdoored model trained without any transparency.
We visualize a set of randomly backdoored samples in ~\autoref{figure:transparentCIFAR}.
As the figure shows, setting the scale to 0.1 makes the triggers hardly visible.

\subsection{Relaxing the Threat Model (Transferability of the Triggers)}
\label{section:relax}

For our dynamic backdoor attacks, we assume the adversary to control the training of the target model.
We now relax this assumption by only allowing them to poison the dataset.

First, it is important to mention that our {\noisyBD} technique does not need to change the training of the target model, i.e., the adversary only needs to poison the training dataset with backdoored images and the corresponding target labels.
Second, for both the {\ban} and {\cban} techniques, the adversary can rely on pre-trained {\ban} and {\cban} models instead of training them jointly with the target model.
In detail, the adversary uses the pre-trained {\ban} and {\cban} model to generate multiple triggers and randomly place them to a set of -- randomly picked -- images.
Then, they poison the training set with this set of backdoored images and their corresponding target labels.

We use the MNIST dataset for evaluation and follow the same target models' structure as previously introduced in~\autoref{section:banEval} and~\autoref{section:cBaNEval}.
However, to show the flexibility of our techniques, we use data from different distributions to pre-train the {\ban} and {\cban} models.
We first use the CIFAR-10 dataset to train backdoored models with the {\ban} (\autoref{section:ban}) and {\cban} (\autoref{section:cban}) techniques.
Next, we use the pre-trained {\ban} and {\cban} models to generate the backdoored dataset and poison the target dataset as previously mentioned.
It is important to mention that the CIFAR-10 based {\ban} and {\cban} models generate 3-channel triggers, to use them to poison the MNIST dataset, we convert them to 1-channel triggers by taking the mean over the different channels.
Finally, we use the poisoned dataset to train the target model.

As expected, the backdoored models achieve a perfect attack success rate (100\%), while keeping the same utility as the backdoored models jointly trained with the {\ban} and {\cban}.
This shows the flexibility of our attacks, i.e., the training procedure can be adapted by the adversary depending on their specific application.
However, it is important to mention that jointly training the models has the advantage of giving the adversary more power, e.g., they can add a customized loss function to the target model while implanting the backdoor.

Finally, as a side-effect of transferring the {\ban} and {\cban}; The poisoning rate for the dynamic backdoor can now be lowered to about 10\%, as there is no joint models trained with the target model anymore.

\subsection{Possible Defenses}

Finally, we propose some possible defenses against our dynamic backdoor attacks.
Intuitively, we use a denoising mechanism to filter triggers (as they can be considered as anomalies/distortions) out of the backdoored inputs.
To this end, we use one of the most common denoising mechanisms, namely autoencoder. 
It works as follows:
First, we train an autoencoder on clean data.
Then, we use this autoencoder to reconstruct/denoise the inputs (by encoding then decoding them).
The noise or triggers in our case are expected to be filtered out of the inputs due to two main reasons.
First, the overfitting of the autoencoders to clean data, and second, the lossy reconstruction process.

To implement our defense, we use the autoencoder to denoise all inputs before forwarding them to the target model.
The autoencoder is expected to remove the trigger from backdoored data, while not significantly changing the clean ones.

To evaluate the efficacy of our proposed defense, we test it against the {\cban} technique, using both the MNIST and CIFAR-10 datasets.
As expected, the backdoored images are not perfectly reconstructed by the autoencoder, i.e., the autoencoder does not fully reconstruct triggers.
Our experiments show that in simple datasets like MNIST, our approach can successfully defend against the backdoor attack, with negligible utility loss (less than 1\%).
However, for more complicated datasets like CIFAR-10, the performance of our defense degrades.
This is due to the high amount of details which hardens the reconstruction process of complex datasets (for both clean and backdoored inputs).
For instance, the accuracy of the target model drops by 4.8\% and 25\% for the clean and backdoored dataset, respectively.

Another possible defense approach is first calculating the distance between the reconstructed input and the original input, then taking the decision to forward the input or not to the model, based on a predetermined threshold.
We plan to explore this approach and other potential methods in future work.

We now discuss another defense, namely data augmentation.
More concretely, we discuss the effect of resizing, cropping, and flipping the target images on our dynamic backdoor attacks.
To this end, we use the CIFAR-10 dataset and test how resizing, cropping, or flipping the backdoored image before querying it to the backdoored model affects the performance, i.e., the ASR and utility.
We evaluate the performance of our simplest and most complex setting, i.e., the random backdoor with a single target label and the cBaN with all possible target labels.
We start with the flipping operation, i.e., we flip each input before querying it to the target model.
Our results show that flipping the inputs reduces the ASR to approximately 88.6\% and 93.4\%, without having a significant effect on the utility for the cBaN and random backdoor, respectively.
This shows that our dynamic backdoor attacks are resilient to flipping.
Second, we test the resizing, i.e., we downsize the input image to 16x16 pixels before scaling it back to 32x32 pixels (the model's expected input size).
Resizing the inputs reduced our ASR to approximately 57.4\% and 66.5\%. However, it also dropped the utility by 15.4\% and 15.9\% for the random backdoor and cBaN, respectively.
This shows that scaling can drop our attack performance by on average 40\%, at the cost of a more than 15\% reduction in utility.
Finally, for cropping, we pad all boarder of the input image by 4 black pixels, i.e., with the value 0, then we randomly select a location to crop the padded image back to its original size (32x32).
Our results show that cropping drops the ASR to about 73.2\% and 89.2\%, while the accuracy drops by 0.7\% and 0.25\% for the cBaN and random backdoor models, respectively.

We next include the three data augmentation techniques in the training of the models and repeat the same experiments, i.e., testing the effect of applying each data augmentation separately at the inference time. 
We observe that the results did not differ significantly from the previous set of experiments; hence we plot the result in the Appendix (\autoref{figure:dataAugTrain}). 

These experiments show that data augmentations can reduce the performance of our dynamic backdoor attacks; however, they cannot prevent it and can drop the utility significantly.
In other words, our attacks are still applicable but with a reduced ASR when applying different data augmentation techniques.

\section{Related Work}
\label{section:relatedwork}

In this section, we discuss some of the related work.
We start with current state-of-the-art backdoor attacks.
Then we discuss the defenses against backdoor attacks,  and finally mention other attacks against machine learning models.

\mypara{Backdoor Attacks}
Gu et al.~\cite{GDG17} introduce BadNets, the first backdoor attack on machine learning models.
BadNets uses the MNIST dataset and a square-like trigger with a fixed location, to show the applicability of the backdoor attacks in the ML settings.
Liu et al.~\cite{LMALZWZ18} 
later propose a more advanced backdooring technique, namely the Trojan attack.
They simplify the threat model of BadNets by eliminating the need for access to the training data used to train the target model.
The Trojan attack reverse-engineers the target model to synthesize training data.
Next, it generates the trigger in a way that maximizes the activation functions of the target model's internal neurons related to the target label.
In other words, the Trojan attack reverse-engineers a trigger and training data to retrain/update the model and implement the backdoor.

The main difference between these two attacks (BadNets and Trojan attacks) and our work is that both attacks only consider static backdoors in terms of triggers' pattern and location.
Our work extends the backdoor attacks to consider dynamic patterns and locations of the triggers.

Nguyen and Tran~\cite{NT20} present an input-aware dynamic backdoor.
Intuitively, they propose a trigger generating network that generates independent triggers for each input, i.e., a new unique trigger is generated for every input.
One main difference between their and our work is the structure of generated triggers.
In our work, we model the triggers to be square-like, while they model it to be scattered pixels/patterns across the image.
One advantage of our triggers is that they can be applied to physical objects/images. For instance, the adversary can print our triggers and attach them to the image/object.
We also show -- in \autoref{section:relax} -- how to relax our threat model to only include poisoning of the training dataset, unlike~\cite{NT20} which requires full access to the training of the target model.
Finally, using our approach, the adversary can generate multiple triggers for the same image, which results in a more flexible attack.
More concretely, if the defender gets access to the backdoored image, the adversary can still trigger the same image using different triggers for the same target label.

We focus on backdoor attacks against image classification models, but backdoor attacks can be extended to other scenarios, such as Federated Learning~\cite{WSRVASLP20}, Video Recognition~\cite{ZMZBCJ20}, Transfer Learning~\cite{YLZZ19}, and Natural Language Processing (NLP)~\cite{CSBMSWZ21}.

To increase the stealthiness of the backdoor, Saha et al.~\cite{SSP20} propose to transform the backdoored images into benign-looking ones, which makes them harder to detect.
Lie et al.~\cite{LMBL20} introduce another approach, namely, the reflection backdoor (Refool), which hides the triggers using mathematical modeling of the physical reflection property.
Another line of research focuses on exploring different methods of implementing backdoors into target models.
Rakin et al.~\cite{RHF20} introduce the Targeted Bit Trojan (TBT) technique, which instead of training the target model, flips some bits in the target models' weights to make it misclassify all the inputs.
Tang et al.~\cite{TDLYH20} present a different approach, where the adversary appends a small Trojan module (TrojanNet) to the target model instead of fully retraining it.

\mypara{Defenses Against Backdoor Attacks} 
Defenses against backdoor attacks  can be classified into model-based defenses and data-based defenses.
First, model-based defenses try to find if a given model contains a backdoor or not.
For instance, Wang et al.~\cite{WYSLVZZ19} propose Neural Cleanse (NC), a backdoor defense method based on reverse engineering.
For each output label, NC tries to generate the smallest trigger, which converts the output of all inputs applied with this trigger to that label.
NC then uses anomaly detection to find if any of the generated triggers are actually a backdoor or not.
Later, Liu et al.~\cite{LLTMAZ19}  propose another model-based defense, namely, ABS.
ABS detects if a target model contains a backdoor or not, by analyzing the behaviour of the target model's inner neurons when introducing different levels of stimulation.
Xu et al. ~\cite{XWLBGL21} present another model-based defense, namely MNTD.
Abstractly, MNTD builds a meta classifier to detect if a given model is backdoored or not.
To this end, it first starts by using a small set of clean data to train multiple shadow clean and backdoored models using different triggers.
Next, it optimizes a set of probing points which is used to query all training models.
Finally, the defender queries each shadow model using the probing points and uses pairs of corresponding predictions and ground-truth labels (i.e., clean or backdoored) to train a meta-classifier. 
The meta classifier takes as input a model then outputs a score for each model, indicating if the model is backdoored or not.

Second, data-based defenses try to find if a given input is clean or backdoored.
For instance, Gao et al.~\cite{GXWCRN19} propose STRIP, a backdoor defense method based on manipulating the input, to find out if it is backdoored or not.
More concretely, STRIP fuses the input with multiple clean data, one at a time.
Then it queries the target model with the generated inputs, and calculates the entropy of the output labels.
Backdoored inputs tend to have lower entropy than clean ones.

Similarly, Doan et al.~\cite{DAR20} presents Februus, which is another data-based defense.
Intuitively, Februus uses GradCAM to find the most contributing regions of the input (with respect to the model's output).
Then it determines if the assigned regions contain a trigger or not.
If it contains a trigger, then Februus removes this part from the input and uses a GAN-based inpainting technique to complete the input again before forwarding it to the model.

\mypara{Attacks Against Machine Learning}
Poisoning attack~\cite{JOBLNL18,SMKID18,BNL12,SBZ22} is another training time attack, in which the adversary manipulates the training data to compromise the target model.
For instance, the adversary can change the ground truth for a subset of the training data to manipulate the decision boundary, or more generally influence the model's behavior.
Shafahi et al.~\cite{SHNSSDG18} further introduce the clean label poisoning attack.
Instead of changing labels, the clean label poisoning attack allows the adversary to modify the training data itself to manipulate the behaviour of the target model.

Another class of ML attacks is the adversarial examples.
Adversarial examples share some similarities with the backdoor attacks.
In this setting, the adversary aims to trick a target classifier into miss classifying a data point by adding controlled noise to it.
Multiple work has explored the privacy and security risks of adversarial examples~\cite{PMGJCS17,VL14,CW17,LV15,TKPGBM17,PMJFCS16,XEQ18}.
Other work explores the adversarial example's potentials in preserving the user's privacy in multiple domains~\cite{OFS17,JG18,ZHRLPB18,JSBZG19}.
The main difference between adversarial examples and backdoor attacks is that backdoor attacks are performed in training time, while adversarial examples are performed after the model is trained and without changing any of the model's parameters.

Beside the above, there are multiple other types of attacks against machine learning models~\cite{LWHSZBCFZ22}, such as membership inference~\cite{SSSS17,HZHBTWB19,CYZF20,LBWBWTGC18,YGFJ18,SS19,SZHBFB19,NSH19,HZ21,HJBGZ21}, model stealing~\cite{TZJRR16,OSF19,WG18,SHHZ22}, model inversion~\cite{FLJLPR14,FJR15,HAP17}, property inference~\cite{GWYGB18,MSCS19,ZCSZ22}, and dataset reconstruction~\cite{SBBFZ20}.

\section{Conclusion}
\label{section:conclusion}

The tremendous progress of machine learning has lead to its adoption in multiple critical real-world applications.
However, it has been shown that ML models are vulnerable to various types of security and privacy attacks.
In this paper, we focus on backdoor attacks where an adversary manipulates the training of the model to intentionally misclassify any input with an added trigger.

Current backdoor attacks only consider static triggers in terms of patterns and locations.
In this work, we propose the first set of dynamic backdoor attacks against deep neural networks (DNN) models, where the trigger can have multiple patterns and locations.
To this end, we propose three different techniques.

Our first technique {\noisyBD} samples triggers from a uniform distribution and places them at random locations of an input.
For the second technique, i.e., Backdoor Generating Network ({\ban}), we propose a novel generative network to construct triggers.
Finally, we introduce conditional Backdoor Generating Network ({\cban}) to generate label specific triggers.

We evaluate our techniques using three benchmark datasets.
The evaluation shows that all our techniques  can achieve almost a perfect backdoor success rate  while preserving the model's utility.
Moreover, we show that our techniques successfully bypass state-of-the-art defense mechanisms against backdoor attacks.

\section*{Acknowledgment}

Ahmed Salem's work was mostly done while at CISPA Helmholtz Center for Information Security.
The research leading to these results has received funding from the European Research Council under the European Union's Seventh Framework Programme (FP7/2007-2013)/ ERC grant agreement no. 610150-imPACT, from the Helmholtz Association within the project ``Trustworthy Federated Data Analytics'' (TFDA) (funding number ZT-I-OO1 4) and from U.S. IARPA TrojAI W911NF-19-S-0012.

\bibliographystyle{plain}
\bibliography{normal_generated_py3}

\appendix
\section{Additional Experimental Results}

\begin{figure}[!hbtp]
\centering
\begin{subfigure}{0.8\columnwidth}
\includegraphics[width=\columnwidth]{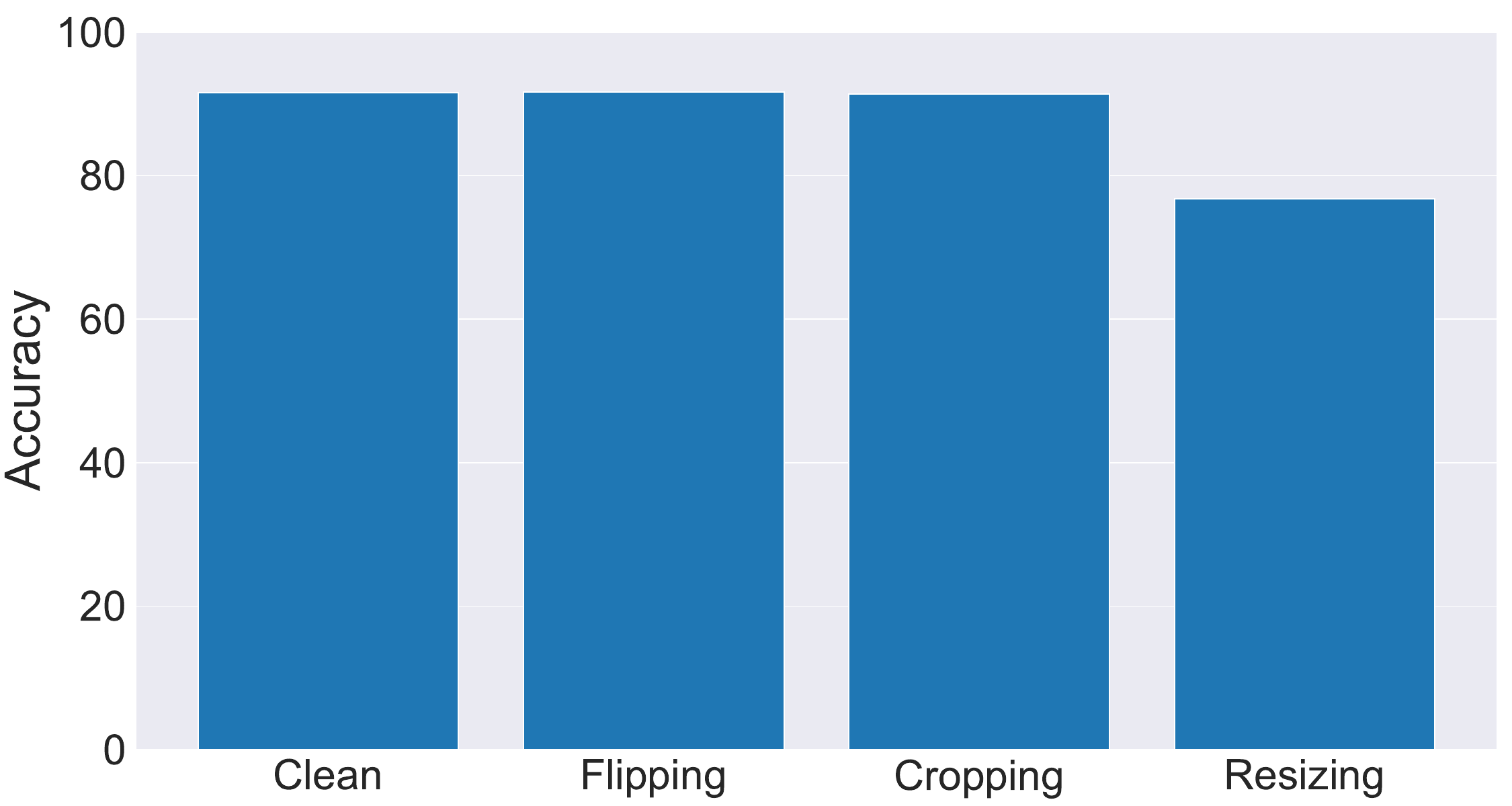}
\caption{Utility}
\label{figure:dataAugTrainUtility}
\end{subfigure}
\begin{subfigure}{0.8\columnwidth}
\includegraphics[width=\columnwidth]{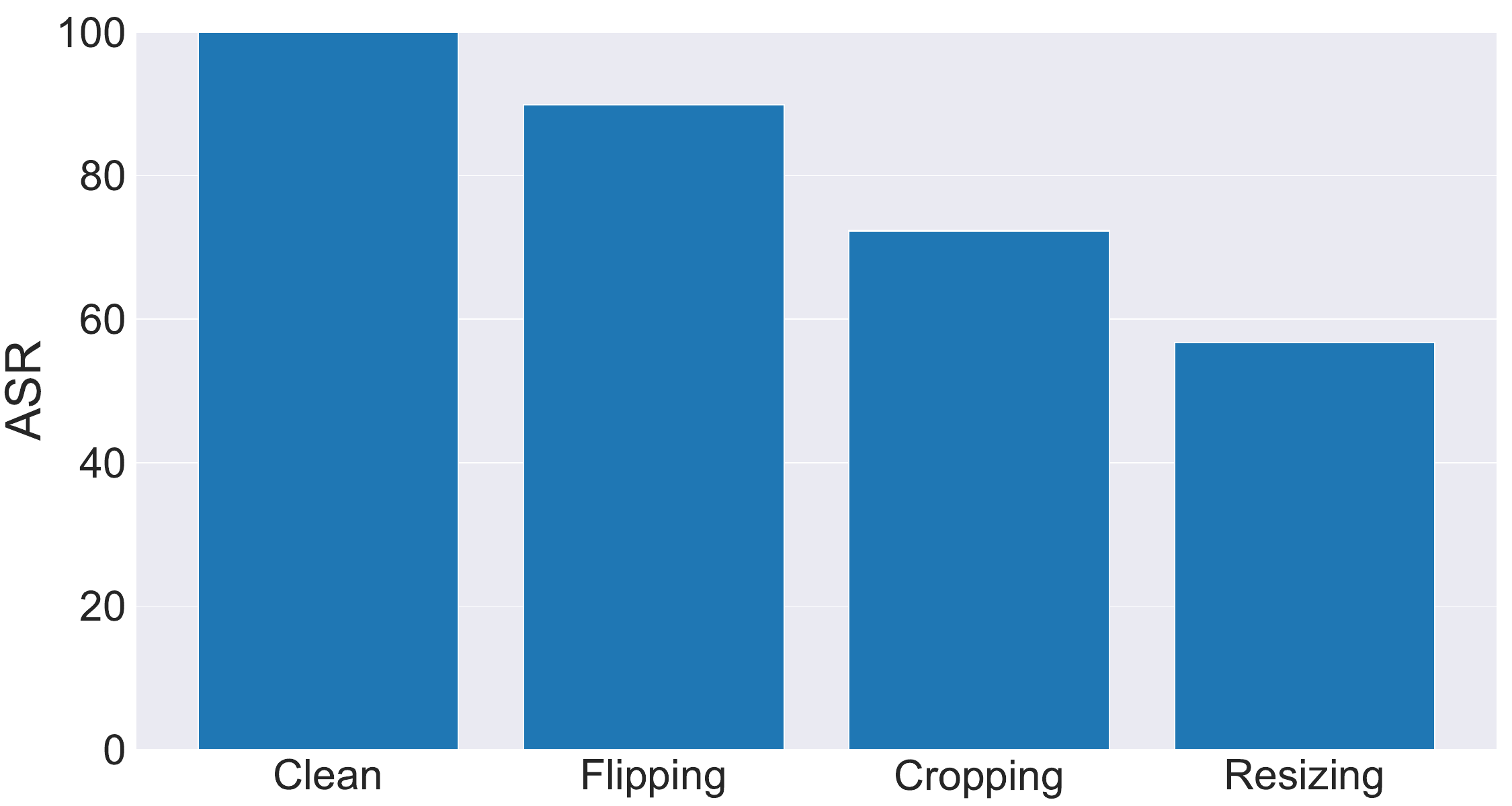}
\caption{ASR}
\label{figure:dataAugTrainASR}
\end{subfigure}
\caption{The performance of the cBaN technique when applying data augmentations techniques when training the target model. \autoref{figure:dataAugTrainUtility} and \autoref{figure:dataAugTrainASR} shows the utility and ASR, respectively.}
\label{figure:dataAugTrain}
\end{figure} 

\end{document}